\documentclass[twocolumn]{aastex631}

\usepackage{natbib}
\usepackage{color}
\usepackage{mathtools}
\usepackage{hyperref} 

\definecolor{mygreen}{rgb}{0.13,0.54,0.13}

\def    \apjl  		{\rm {ApJL}}

\def    \apjl  		{\rm {ApJL}}

\def	\cm		{\,{\rm {cm}}}
\def	\K		{\,{\rm K}}
\def	\g		{\,{\rm {g}}}
\def	\mum	{\,{\mu \rm{m}}}

\def \bea {\begin{eqnarray}}
\def \ena {\end{eqnarray}}

\def 	\bE	{\boldsymbol{E}}

\def	\bJ	{\boldsymbol{J}} 
\def	\bk	{\boldsymbol{k}}

\def    \bmu    {{\hbox{\boldsym\char'026}}}	

\def	\bp	{\boldsymbol{p}}

\def	\bv	{\boldsymbol{v}}

\def	\C	{{\rm C}}

\def	\cm	{\,{\rm cm}}
\def	\km	{\,{\rm km}}

\def	\max	{\,{\rm max}}
\def	\d	{{\rm d}}

\def	\D	{{\rm D}}

\def	\erg	{\,{\rm erg}}

\def	\g	{\,{\rm g}}
\def	\gas	{\,{\rm gas}}
\def	\G	{{\rm G}}

\def	\H	{{\rm H}}

\def	\N	{{\rm N}}

\def	\s	{\,{\rm s}}

\def	\V	{{\rm V}}
\def	\Bar	{{\rm Bar}}
\def	\rad	{{\rm rad}}
\def	\yr	{\,{\rm yr}}

\def \St {{\rm St}}

\def	\ahat		{\hat{\bf a}}

\def    \Bv     	{\boldsymbol{B}}

\def    \kv     	{\boldsymbol{k}}

\def    \gas     	{{\rm gas}}

\font\mib=cmmib10

\def\bOmega{\hbox{\mib\char"0A}}
\def\bmu{\hbox{\mib\char"16}}

\renewcommand{\arraystretch}{1.2}

\begin{document}
\shorttitle{Alignment of carbonaceous dust}
\shortauthors{Hoang, Phan and Tram}
\title{Internal and external alignment of carbonaceous grains within the radiative torque paradigm}

\author{Thiem Hoang}
\affiliation{Korea Astronomy and Space Science Institute, Daejeon 34055, Republic of Korea, \href{mailto:thiemhoang@kasi.re.kr}{thiemhoang@kasi.re.kr}}
\affiliation{Korea University of Science and Technology, 217 Gajeong-ro, Yuseong-gu, Daejeon, 34113, Republic of Korea}

\author{Vo Hong Minh Phan}
\affiliation{Institute for Theoretical Particle Physics and Cosmology (TTK), RWTH Aachen University, 52056 Aachen, Germany, \href{mailto:vhmphan@physik.rwth-aachen.de}{vhmphan@physik.rwth-aachen.de}}

\author{Le Ngoc Tram}
\affiliation{Max-Planck-Institut f\"ur Radioastronomie, Auf dem H\"ugel 69, 53121, Bonn, Germany, \href{mailto:nle@mpifr-bonn.mpg.de}{nle@mpifr-bonn.mpg.de}}

\begin{abstract}
We study the internal and external alignment of carbonaceous grains, including graphite and hydrogenated amorphous carbon (HAC),  in the interstellar medium (ISM) within the RAdiative Torque (RAT) paradigm. For internal alignment (IA), we find that HAC grains having nuclear paramagnetism due to hydrogen protons can have efficient nuclear relaxation, whereas both HAC and graphite grains can have efficient inelastic relaxation for grains aligned both at low$-J$ and high$-J$ attractors. For external alignment, HAC and graphite grains can align with the radiation direction ($k$-RAT) at low$-J$ attractors but cannot have stable alignment at high$-J$ attractors due to the suppression of radiative precession. HAC also has slow Larmor precession compared to the randomization by gas collisions and cannot be aligned with the magnetic field ($B$-RAT). Small HAC grains of $a<0.05\mum$ drifting through the diffuse ISM can be weakly aligned along the induced electric field ($E$-RAT) at high$-J$ attractors due to its fast precession. Paramagnetic relaxation by nuclear magnetism is found inefficient for HAC grains due to the rapid suppression of nuclear susceptibility when grains rotate at high$-J$ attractors. We then study the alignment of carbon dust in the envelope of a typical C-rich Asymptotic Giant Branch star, IRC+10216. We find that grains aligned at low$-J$ attractors can occur via $k$-RAT with the wrong IA in the inner region but via $B$-RAT in the outer region. However, grains aligned at high$-J$ attractors have the right IA alignment via $k$-RAT due to efficient inelastic relaxation. The polarization pattern observed toward IRC+10216 by SOFIA/HAWC+ can reproduced when only grains at low$-J$ attractors are present due to removal of grains at high$-J$ attractors by the RAT disruption.
\end{abstract}

\section{Introduction}
Dust and magnetic field are ubiquitous in the universe and play important roles in astrophysics. The polarization of distant starlight (\citealt{Hall:1949p5890}; \citealt{Hiltner:1949p5851}) and polarized thermal dust emission \citep{Hildebrand:1988p2566} reveal that interstellar dust grains are non-spherical and aligned with the interstellar magnetic field. As a result, dust polarization is widely used to map magnetic fields from the diffuse interstellar medium (ISM) to star-forming regions \citep{Crutcher:2010p318,2019FrASS...6...15P}. Spectropolarimetric observations show the strongly polarized spectral feature of silicate at $10\mum$ \citep{2000MNRAS.312..327S}, revealing that silicate grains are well aligned. The physics of grain alignment for silicate grains has been recently established, with RAdiative Torque (RAT) alignment as a leading mechanism (see \citealt{LAH15,2015ARA&A..53..501A} for reviews).

Silicate dust grains containing unpaired electrons, which are paramagnetic material, possess a magnetic moment primarily due to the grain rotation via the Barnett effect \citep{Barnett:1915p6353}. The magnetic torque induced by the interaction of the grain magnetic moment with the ambient magnetic field causes the Larmor precession of the grain angular momentum ($\bJ$) around $\Bv$, which couples the grain and the magnetic field. In general, the process of grain alignment with a preferred direction in space (e.g., $\Bv$) includes (1) the alignment of the grain axis of major inertia ($\ahat_{1}$) with the angular momentum (so-called internal alignment) and (2) the alignment of $\bJ$ with the magnetic field ($\Bv$; so-called external alignment). The internal alignment of paramagnetic grains can be induced by the Barnett relaxation and inelastic relaxation effects \citep{1979ApJ...231..404P}. 

The external alignment used to be thought to arise from paramagnetic relaxation \citep{1951ApJ...114..206D}, namely D-G), which occurs for paramagnetic grains. Yet, modern understanding of grain alignment establishes RATs (\citealt{1976Ap&SS..43..291D}; \citealt{1996ApJ...470..551D}; \citealt{2007MNRAS.378..910L}) as a leading mechanism for grain alignment (\citealt{1997ApJ...480..633D}; \citealt{Hoang:2008gb,2016ApJ...831..159H}). Following the RAT alignment theory, RATs can spin up the grains to suprathermal rotation (i.e., with the grain angular momentum greater than its thermal value) and align them with the magnetic field. Moreover, silicate grains having embedded iron inclusions can achieve perfect alignment thanks to the joint action of enhanced magnetic relaxation and suprathermal rotation by RATs, which is known as the MRAT mechanism \citep{2016ApJ...831..159H}. The RAT alignment theory is observationally tested and becomes the leading mechanism for describing grain alignment and interpreting dust polarization (see \citealt{2015ARA&A..53..501A} and \citealt{LAH15} for reviews). The latest unified model of interstellar dust by \cite{Hensley.2022eg} requires perfect alignment of large grains, which is only achieved by the MRAT mechanism \citep{2016ApJ...831..159H}.

Carbonaceous grains are considered diamagnetic due to the lack of unpaired electrons. Thus, they are expected not to be aligned with the $B$-field. The non-detection of interstellar polarization of the 3.4 $\mum$ C-H feature \citep{2006ApJ...651..268C} also implies that carbonaceous grains are not aligned or have a spherical shape. However, the question of whether carbonaceous grains are aligned is still unclear. In the ISM, grains of different compositions may be mixed together due to continuous collisions, which is difficult to constrain the alignment of carbon grains. However, in environments of newly formed dust, such as the envelope of Asymptotic Giant Branch (AGB) stars or supernova ejecta, silicate and carbon dust components are likely to present in separate forms. Recently, polarized thermal dust emission has been detected from such environments, including the circumstellar envelope around the carbon-rich AGB star, IRC +10216 \citep{Andersson.2022} and the ejecta of core-collapse supernovae \citep{Rho.2022,Chastenet.2022}. Therefore, a detailed study for alignment of carbon dust is crucial to interpret the dust polarization and constrain dust composition.

Note that, although carbon dust is the major dust component of the ISM, the exact form of carbon solid is still hotly debated. The traditional interstellar dust model assumes the graphite form for carbon dust \citep{1984ApJ...285...89D}. Although pure carbon dust like graphite can be formed in the envelope of C-rich AGB stars, they are bombarded by interstellar H atoms when being injected into the ISM. Thus, the hydrogenated amorphous carbon (HAC) dust is previously thought to be an important carbon solid in the ISM \citep{2008A&A...492..127S}. In the model by \citet{2013A&A...558A..62J}, most of C is in the form of HAC, and only a small fraction is in the mantle covering the silicate core. \citet{Furton.1999} estimated that the H fraction to C is $f_{\rm H/C}\sim 0.5$, so that the presence of hydrogen protons makes HAC become a paramagnetic material \citep{Purcell:1969p3641}. 

\cite{1999ApJ...520L..67L} suggested that nuclear relaxation, which is an effect similar to Barnett relaxation but induced by nuclear magnetism, can produce efficient internal alignment for carbon dust with H attachment (e.g., HAC). Moreover, although graphite grains do not exhibit magnetism, their internal alignment can be caused by inelastic relaxation \citep{1979ApJ...231..404P,1999MNRAS.303..673L}.

The modern understanding of grain alignment physics based on RATs (aka. {\it RAT alignment paradigm}) establishes that grains can only be efficiently aligned if they spin suprathermally by RATs \citep{2016ApJ...831..159H,2016ApJ...821...91H}. Interestingly, analytical model \citep{2007MNRAS.378..910L} and numerical calculations \citep{2007MNRAS.378..910L,Herranen.2021} show a weak dependence of RATs on the grain composition. Therefore, we expect that carbonaceous grains can spin suprathermally and be stably aligned by RATs. At the same time, the spin-down effect of RATs induces the grain alignment at thermal rotation (so-called low$-J$ attractors). As a result, the combination of internal alignment by nuclear relaxation in HAC or inelastic relaxation (in both HAC and graphite) with the spin-up by RATs perhaps induces efficient alignment of carbonaceous grains.

More recently, \cite{2019ApJ...883..122L} discussed the effect of nuclear magnetism for external alignment of $\bJ$ with $\Bv$ via the magnetic relaxation \citep{1951ApJ...114..206D} and noticed the suppression of relaxation at a high angular velocity of $\omega>2/\tau_{n}\sim 10^{4}$ rad/s with $\tau_{n}$ the nuclear spin-spin relaxation timescale. \cite{Lazarian.2020bwy} also discussed the alignment of carbonaceous grains by RATs along the electric field (${\bf E}$) induced by the grain drifting at velocity $v_{d}$ relative to the ambient magnetic field (i.e., ${\bf E}=[{\bf v}_{d}/c\times \Bv]$), which is called the $E$-RAT alignment. The author noted the uncertainty on $E$-RAT alignment because of the dependence of the electric precession rate on the grain angular momentum. Previously, a detailed study of grain alignment with the electric field was presented in \cite{2014MNRAS.438..680H}.

Interestingly, observations by SOFIA/HAWC+ toward a C-rich star, IRC +10216, reported the polarization of thermal dust emission with a radial polarization pattern along the radial direction \citep{Andersson.2022}. The authors explain such a radial polarization pattern due to $k$-RAT alignment with inefficient but wrong internal alignment (IA) because of slow internal relaxation. We note that the wrong IA is only possible when grains have slow internal relaxation compared to the grain randomization by gas collisions \citep{2009ApJ...695.1457H}, but inelastic relaxation is expected to be efficient for small dust grains \citep{1979ApJ...231..404P,1999MNRAS.303..673L}. This observation, combined with the non-detection of polarized $3.4\mum$ C-H feature, appears to challenge the expectation from the RAT alignment theory. Therefore, the main goal of this paper is (1) to study whether carbonaceous grains can be aligned within the RAT paradigm and (2) to apply the obtained results for understanding thermal dust polarization toward C-rich AGB stars, such as IRC +10216.

The rest of the paper is organized as follows. In Section \ref{sec:magnetic}, we describe the grain model, nuclear magnetism due to hydrogen nuclear spins, and resulting magnetic moments for HAC dust. In Section \ref{sec:RATparadigm}, we summarize the essential components of the RAT alignment paradigm. In Section \ref{sec:internal} we present the effects of nuclear relaxation and inelastic relaxation for internal alignment within suprathermally rotating grains by RATs. In Section \ref{sec:external}, we discuss in detail the external alignment of carbon dust along the radiation direction, the magnetic field, and the electric field for the general conditions of the ISM. We then apply our theoretical analysis for studying the alignment of carbon dust in the envelope of C-rich AGB stars in Section \ref{sec:AGB}. An extended discussion is presented in Section \ref{sec:discuss}. A summary of our results is presented in Section \ref{sec:concl}.

\section{Grain Model and Magnetic Properties}\label{sec:magnetic}
\subsection{Grain model and assumptions}
\begin{figure}
\includegraphics[width=0.45\textwidth]{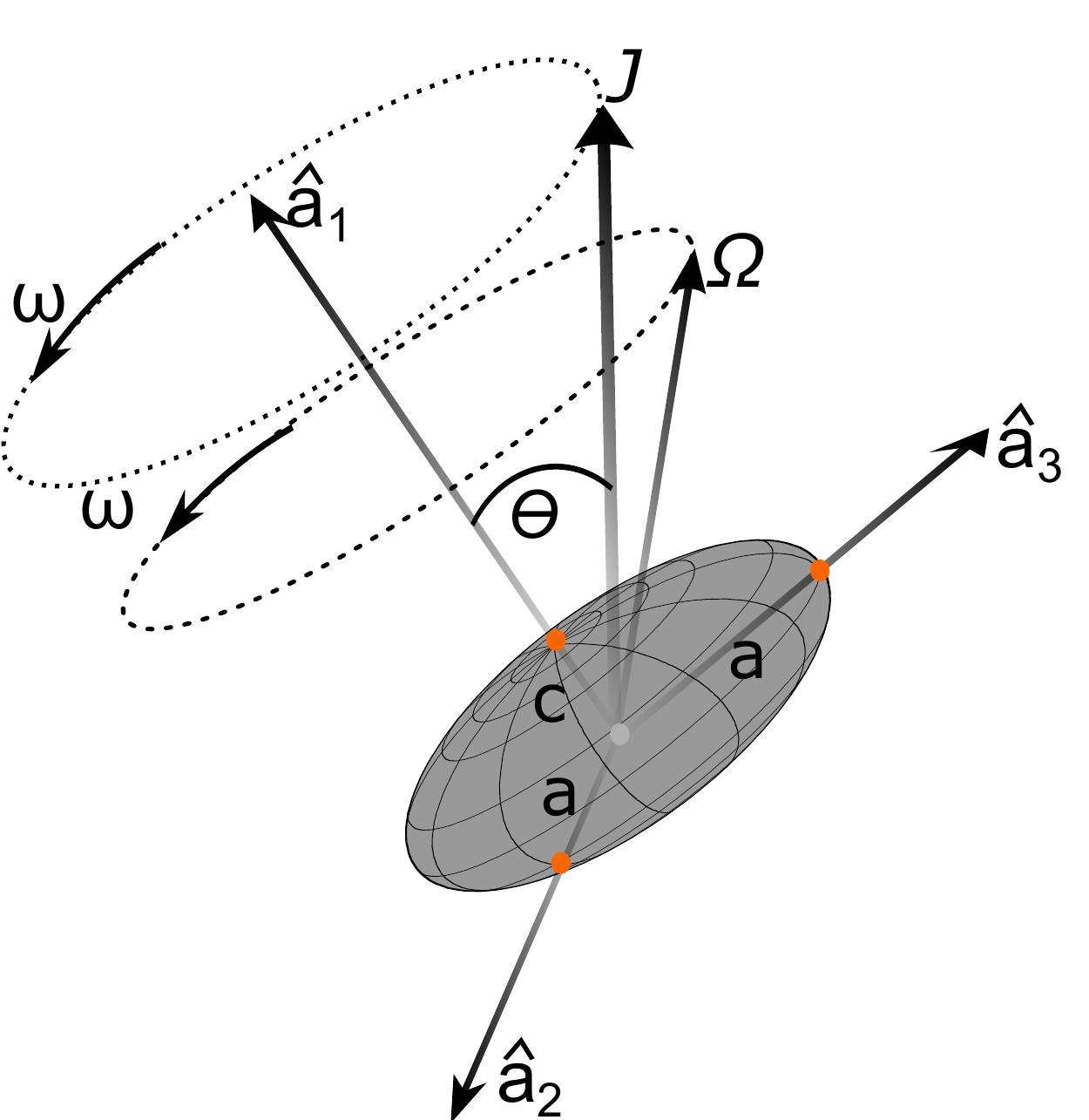}
\caption{Illustration of torque-free motion of a spheroidal grain of principal axes $\ahat_{1}\ahat_{2}\ahat_{3}$. In the body frame, the angular momentum ($\bJ$) and angular velocity ($\bOmega$) are both precessing around the axis of maximum inertia $\ahat_{1}$ with the angular rate $\omega$. Efficient internal relaxation induces the alignment of both $\bJ$ and $\bOmega$ with $\ahat_{1}$ (internal alignment). From \cite{Hoang.20220o}.}
\label{fig:torque-free}
\end{figure}

Here we follow the grain model as in \cite{Hoang.20220o}. A dust grain of triaxial (irregular) shape is described by the principal axes $\ahat_{1},\ahat_{2},\ahat_{3}$. Let $I_{1}> I_{2}\ge I_{3}$ be the principal moments of inertia along the principal axes, respectively. For numerical estimates, we assume an oblate spheroidal grain with $I_{1}> I_{2}=I_{3}$. Let us denote $I_{\|}\equiv I_{1}$ and $I_{\perp}=I_{2}$ for simplicity. The length of the symmetry (semi-minor) axis is denoted by $c$ and the lengths of the semi-major axes are denoted by $a$ and $b$ with $a=b$ (see Figure \ref{fig:torque-free}).

The principal moments of inertia for the rotation parallel and perpendicular to the grain symmetry axis are given by
\bea
I_{\|}&=&\frac{8\pi}{15}\rho a^{4}c=\frac{8\pi}{15}\rho sa^{5},\label{eq:Ipar}\\
I_{\perp}&=&\frac{4\pi}{15}\rho a^{2}c\left(a^{2}+c^{2}\right)=\frac{4\pi}{15}\rho sa^{5}\left(1+s^{2}\right),\label{eq:Iparperp}
\ena 
where $\rho$ is the mass density of the grain with $\rho\approx 2.2\g\cm^{-3}$ for carbonaceous grains, and $s=c/a<1$ is the axial ratio. The ratio of the principal inertia moments is $h=I_{\|}/I_{\perp}=2/(1+s^{2})>1$ for $s<1$ (see also \citealt{2014MNRAS.438..680H}). The effective size of the equivalent sphere of the same volume as the oblate spheroid is $a_{\rm eff}=s^{1/3}a$.

In a general case, the grain angular momentum ($\bJ$) and angular velocity ($\bOmega$) are not parallel to the axis of maximum inertia, $\ahat_{1}$. Let $\theta$ be the angle between $\bJ$ and $\ahat_{1}$. Thus, $\Omega_{1}=J_{1}/I_{1}=J\cos\theta/I_{1}\equiv \Omega_{0}\cos\theta$ with $\Omega_{0}=J/I_{1}$. The precession of $\bJ$ and $\bOmega$ around $\ahat_{1}$ occurs at an angular rate of $\omega=(h-1)\Omega_{1}=(h-1)\Omega_{0}\cos\theta$, which is evaluated at $\theta=\pi/4$ for numerical estimates (e.g., \citealt{1979ApJ...231..404P}; \citealt{Hoang:2010jy}).

In the absence of radiation-dust interaction, gas-grain collisions can establish the thermal equilibrium for the grain rotation energy. The thermal rotation energy of grains is then defined by the gas kinetic energy, such that $\Omega_{T}=(kT_{\rm gas}/I_{\|})^{1/2}=(kT_{\rm gas}/I_{\|})^{1/2}=1.65\times 10^{4}\hat{\rho}^{-1/2}T_{2}^{1/2}s^{-1/2}a_{-5}^{-5/2}\rm rad\s^{-1}$ with $T_{2}=T_{\gas}/100\K$. The grain thermal angular momentum is $J_{T}=I_{\|}\Omega_{T}$.

The grain rotation can be damped by various processes, including sticky collisions of gas species followed by evaporation and infrared emission (see e.g., \citealt{Hoang.2021}). The rotational damping time due to gas collisions is given by
\bea
\tau_{\gas}&=&\frac{3}{4\sqrt{\pi}}\frac{I_{\|}}{1.2n_{\rm H}m_{\rm H}
v_{\rm T}a^{4}\Gamma_{\|}}\nonumber\\
&\simeq& 2.62\times 10^{5}\hat{\rho}sa_{-5}\left(\frac{1}{n_{1}T_{2}^{1/2}\Gamma_{\|}}\right)~{\rm yr},\label{eq:tgas}
\ena
where $a_{-5}=a/(10^{-5}\cm)$, $\hat{\rho}=\rho/(3\g\cm^{-3})$, and $\Gamma_{\|}$ is the geometrical factor of unity order (\citealt{1993ApJ...418..287R}; \citealt{2009ApJ...695.1457H}). The gas density and temperature are normalized to its typical value of a diffuse ISM with $n_{1}=n_{\H}/(10^{1}\,\cm^{-3})$ and $T_{2}=T_{\rm gas}/(100\,{\rm K})$. The axial ratio has been chosen to be $s=1/2$, which is used throughout this paper.

Subject to a radiation field of radiation energy density $u_{\rm rad}$, dust grains are heated to high temperatures and quickly cool down by infrared emission. Infrared emission also results in the damping of the grain rotation because emitting photons carry away some grain angular momentum. For large grains that can have thermal equilibrium at a temperature $T_{d}$, the IR damping rate $\tau_{\rm IR}^{-1}=F_{\rm IR}\tau^{-1}_{\gas}$ with $F_{\rm IR}$ being the dimensionless IR damping parameter (see \citealt{1998ApJ...508..157D}), given by
\bea
F_{\rm IR}\simeq \left(\frac{1.2}{a_{-5}}\right)\left(\frac{U^{2/3}}{n_{1}T_{2}^{1/2}}\right),\label{eq:FIR}
\ena 
where $U=u_{\rm rad}/u_{\rm MMP83}$ be the strength of the radiation field where $u_{\rm MMP83}$ is the energy density of the radiation field in the solar neighborhood from \cite{1983A&A...128..212M}.

\subsection{Nuclear magnetic susceptibility of HAC}
The spin of nuclei, such as hydrogen protons, can induce nuclear magnetic moment with $\mu_{n}=g_{n}\mu_{N}$ where $g_{n}\simeq 3.1$ is the nuclear $g$-factor and $\mu_{N}=e\hbar/2m_{pro}c=5.05 \times10^{-24}$ erg G$^{-1}$ is the proton magneton. The zero-frequency susceptibility of proton nuclei is given by (see \citealt{1999ApJ...520L..67L})
\bea
\chi_{n}(0)&=&\left(\frac{n_{n}\mu_{n}^{2}}{3kT_{d}}\right),\nonumber\\
&=&6.2\times 10^{-12}\left(\frac{\mu_{n}}{\mu_{N}}\right)^{2}\left(\frac{n_{n}}{10^{22}\cm^{-3}}\right)\left(\frac{100\K}{T_{d}}\right),\label{eq:chi_n}
\ena
where $n_{n}$ is the number density of hydrogen protons in HAC. 

If the fraction of hydrogen protons is too low $n_n\ll 10^{22}$ cm$^{-3}$, i.e., the fraction of hydrogen protons, $f_{\rm prot}=n_{n}/n\sim 10\%$,\footnote{The atomic number density of carbon grain is $n=\rho/m_{C}\approx 10^{23}\cm^{-3}$ with $\rho\approx 2.2\g\cm^{-3}$.} the nuclei of $^{13}$C might become dominant for the magnetization of carbonaceous grains and the zero-frequency susceptibility. In this case, the susceptibility could also be obtained using Eq. \eqref{eq:chi_n} assuming $g_n=\mu_n/\mu_N\simeq 0.8$. This might be the case in the envelope of C-rich AGB stars where grains are believed to be mostly in form of graphite. The effect of different composition on the nuclear magnetism (hydrogen in HAC or $^{13}$C dominated grains) will be discussed in more detail in Section \ref{sec:AGB}. However, we will focus mostly on HAC grains with $g_n=3.1$ and $n_n=10^{22}\cm^{-3}$ (unless otherwise stated) for purposes of discussion in other sections of this paper.  


The imaginary part of the complex magnetic susceptibility is a function of the grain rotation frequency and is usually represented as $\chi_{2}(\omega)=\omega K(\omega)$ where $K(\omega)$ is the function obtained from the magnetization dynamics equation. The frequency-dependence nuclear magnetic susceptibility is given by
\bea
K_{n}(\omega) &=& \frac{\chi_{n}(0)\tau_{n}}{[1+(\omega/\omega_{\rm damp})^{2}]^{2}},\label{eq:Kn}
\ena
where $\omega_{\rm damp}=2/\tau_{n}$ is the damping frequency with $\tau_{n}$ the time of nuclear spin-spin relaxation given by
\bea
\tau_{n}^{-1}=\tau_{ne}^{-1}+\tau_{nn}^{-1},
\ena
where 
\bea
\tau_{ne}\approx 2.3\times 10^{-4}(3.1/g_{n})^{2}(10^{22}\cm^{-3}/n_{e})\,\s
\ena
with $n_e$ and $n_n$ the number densities of electrons and nucleons in the dust, respectively, and $\tau_{nn}\approx 0.58\tau_{ne} (n_{e}/n_{n})$ (\citealt{1999ApJ...520L..67L}; \citealt{2019ApJ...883..122L}). Here we consider the different values of $n_{n}$ to account for the different fractions of H atoms in HAC.

Equation (\ref{eq:Kn}) implies that, for slow rotation of $\omega<\omega_{\rm damp}, $ $K_{n} =\chi_{n}(0)\tau_{n}\propto n_{n}\mu_{n}^{2}\times (1/n_{n}g_{n}\mu_{N})\propto \mu_{n}^{2}$, which does not depend on the density of unpaired protons. However, for faster rotation, the suppression term, $1+(\omega/\omega_{\rm damp})^{2}$ becomes important, and $K_{n}\sim \chi_{n}(0)/\tau_{n}\sim n_{n}\chi_{n}(0)$. 

\begin{figure}
\includegraphics[width=0.5\textwidth]{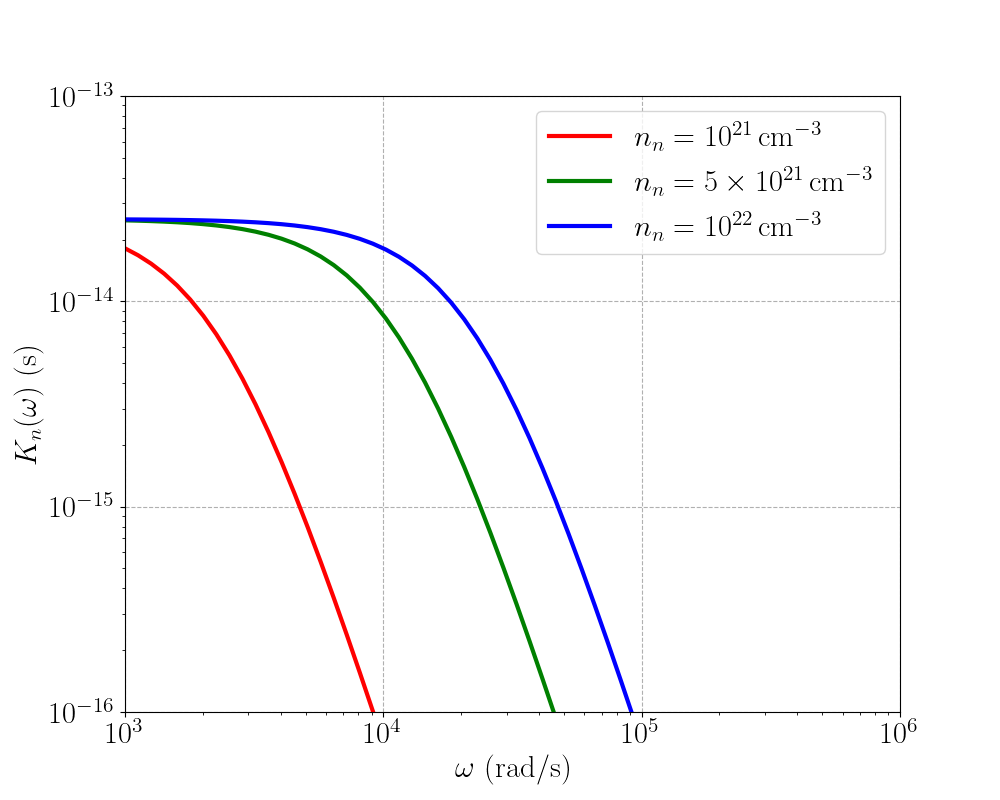}
\caption{Frequency-dependence susceptibility for different values of the nucleon number density $n_n=10^{21}$ cm$^{-3}$ (red), $n_n=5\times 10^{21}$ cm$^{-3}$ (green), and $n_n=10^{22}$ cm$^{-3}$ (blue). A rapid decrease of the susceptibility with increasing frequency is seen.}
\label{fig:Komega}
\end{figure}

Figure \ref{fig:Komega} shows $K_{n}(\omega)$ for different rotation frequency $\omega$. The value of $K$ is almost flat (blue and green lines) and then decreases rapidly with the frequency for $\omega\gtrsim \omega_{\rm damp}\sim 10^{4}$ rad/s. 


\subsection{Grain magnetic moment}
A rotating paramagnetic HAC grain can acquire a magnetic moment thanks to the Barnett effect (see \citealt{Barnett:1915p6353}; \citealt{1976Ap&SS..43..257D}) and the rotation of its charged body (\citealt{1972MNRAS.158...63M}). The Barnett effect, which is shown to be much stronger than the latter, induces a magnetic moment proportional to the grain angular velocity ($\bOmega$):
\bea
\bmu_{\Bar}=\frac{\chi_{n}(0)V}{\gamma_{n}}\bOmega=\frac{\chi_{n}(0)V\hbar}{g_{n}\mu_{N}}\bOmega,\label{eq:muBar}
\ena
where $V=4\pi sa^{3}/3$ is the grain volume, $\chi_{n}(0)$ is the nuclear magnetic susceptibility of the grain at rest given by Equation \eqref{eq:chi_n}.

\subsection{Grain charge and electric dipole moment}\label{sec:charge}
Dust grains in the ISM can get charged by collisions with electrons and protons in the gas and photoemission by UV photons. The equilibrium charge for the grain of size $a$ induces an electric potential:
\bea
\phi&=&\frac{Q}{a}\approx 4.79\times 10^{-5}\langle Z\rangle a_{-5}^{-1} \frac{\rm statC}{\cm}\nonumber\\
&\approx& 0.014\langle Z\rangle a_{-5}^{-1}{\rm V},\label{eq:phi}
\ena
where with $Q=e\langle Z\rangle$ the grain charge and mean charge number $\langle Z\rangle$. 

The charge distribution on the grain surface is likely asymmetric, with the charge center off the center of mass. This induces an electric dipole moment of a magnitude
\bea
p_{Z}= (\epsilon a)Q=10^{-15} \epsilon_{-2}a_{-5}^{2}\left(\frac{|\phi|}{0.3\rm V}\right)~ {\rm statC} \cm,\label{eq:pZ}
\ena
where $\epsilon a$ is the displacement of the charge center from the center of mass and $\epsilon_{-2}=\epsilon/0.01$ \citep{2014MNRAS.438..680H}.

The net charge of a grain induced by collisions and photoelectric effect depends on the UV radiation strength ($G$) and electron density ($n_{e}$) as $\phi\propto G\sqrt{T_{\gas}}/n_{e}$ (see \citealt{2011piim.book.....D}. In cold and dense regions, grain charging is dominated by collisions with electrons and ions in the gas because of the lack of stellar UV photons. The equilibrium grain charge by collisional charging is $\phi= -2.504(kT_{\gas}/e)\simeq -0.0216T_{2} {\rm V}$ or $\langle Z\rangle=\phi a/e=-1.5a_{-5}T_{2}$, which does not depend on the gas ionization fraction. 

\section{The RAdiative Torque (RAT) Alignment  Paradigm}\label{sec:RATparadigm}
Here we describe the essential elements of the RAT alignment paradigm, including the spin-up, precession, and alignment, which will be used to quantify the internal and external of carbonaceous grains. We follow the same structure as in \cite{Hoang.20220o}, which studied the alignment of very large grains with iron inclusions in protostellar environments. 

\subsection{Radiative torques and fundamental effects}

The interaction of an anisotropic radiation field with a helical grain can induce RATs due to the differential scattering/absorption of left- and right-handed circularly polarized photons \citep{1976Ap&SS..43..291D}. Latter, RATs were numerically demonstrated in \cite{1996ApJ...470..551D} for three irregular grain shapes. \cite{2007MNRAS.378..910L} introduced an Analytical Model (AMO) of RATs which is based on a helical grain consisting of an oblate spheroid and a weightless mirror. The AMO is shown to reproduce the basic properties of RATs obtained from numerical calculations for realistically irregular grain shapes \citep{2007MNRAS.378..910L,Hoang:2008gb,Herranen.2021}, and enables us to make quantitative predictions for various conditions \citep{2014MNRAS.438..680H} and dust compositions \citep{2008ApJ...676L..25L,2009ApJ...695.1457H,2016ApJ...831..159H}. Many predictions were observationally tested (see \citealt{2015ARA&A..53..501A}). In particular, RATs are found to depend weakly on the grain composition (carbonaceous vs. silicate material).

\begin{figure}
\includegraphics[width=0.5\textwidth]{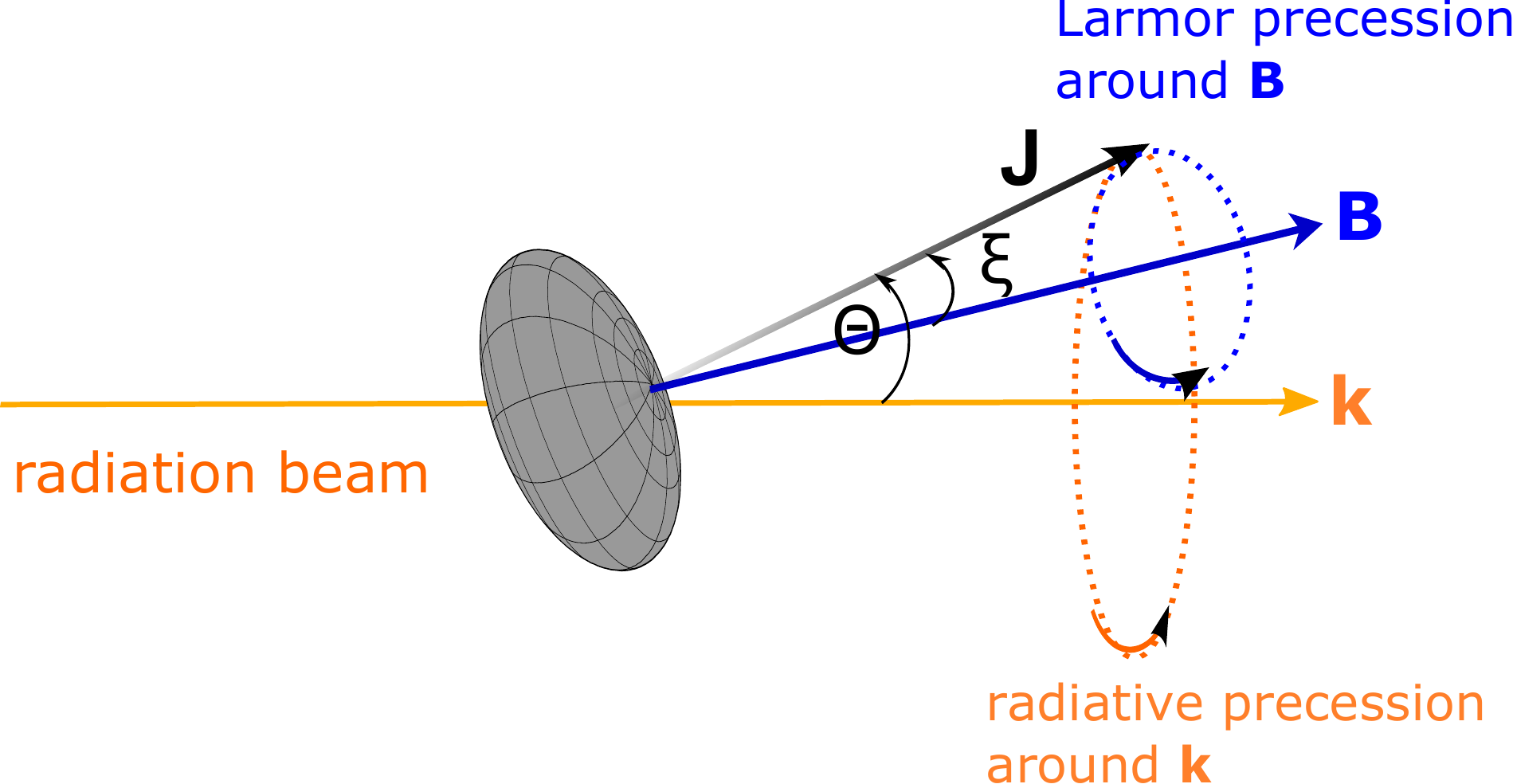}
\caption{Schematic illustration of the precession of the grain angular momentum (${\bf J}$) around the radiation direction (${\bf k}$) and the magnetic field (${\bf B}$), assuming the perfect internal alignment of the grain shortest axis with ${\bf J}$. The grain magnetic moment is directed along ${\bf J}$ due to the Barnett effect. The fastest precession establishes the axis of grain alignment by RATs. From \cite{Hoang.20220o}.}
\label{fig:kB_RAT}
\end{figure}

Previous studies \citep{1997ApJ...480..633D,2007MNRAS.378..910L,Hoang:2008gb} show that RATs in general can induce three important effects: (1) the spin-up to suprathermal rotation as well as spin-down to thermal rotation, (2) the grain precession around the radiation direction, and (3) the alignment of $\bJ$ along the radiation $\kv$ (see Figure \ref{fig:kB_RAT}).

\subsection{Spin-up and suprathermal rotation}

Following \cite{Hoang.2021}, the maximum angular velocity of grains spun-up by RATs is given by
 \bea
\Omega_{\rm RAT}&=& \frac{3\gamma u_{\rm rad}a_{\rm eff}\bar{\lambda}^{-2}}{1.6n_{\rm H}\sqrt{2\pi m_{\rm H}kT_{\rm gas}}}\left(\frac{1}{1+F_{\rm IR}}\right)\nonumber\\
&\simeq &2.97\times 10^{8} s^{1/3}a_{-5}\left(\frac{\bar{\lambda}}{1.2\mum}\right)^{-2}
\left(\frac{\gamma U}{n_{1}T_{2}^{1/2}}\right)\nonumber\\
&&\times\left(\frac{1}{1+F_{\rm IR}}\right)\rad\s^{-1},\label{eq:omega_RAT1}
\ena
for grains with $a_{\rm eff}\lesssim a_{\rm trans}\simeq \bar{\lambda}/2.5$, and
\bea
\Omega_{\rm RAT}&=&\frac{1.5\gamma u_{\rm rad}\bar{\lambda}a_{\rm eff}^{-2}}{16n_{\rm H}\sqrt{2\pi m_{\rm H}kT_{\rm gas}}}\left(\frac{1}{1+F_{\rm IR}}\right)\nonumber\\
&\simeq& 2.56\times 10^{10}s^{-2/3}a_{-5}^{-2}\left(\frac{\bar{\lambda}}{1.2\mum}\right) \left(\frac{\gamma U}{n_{1}T_{2}^{1/2}}\right)\nonumber\\
&&\times\left(\frac{1}{1+F_{\rm IR}}\right)\rad\s^{-1},\label{eq:omega_RAT2}
\ena
for grains with $a_{\rm eff}>a_{\rm trans}$. Here $a_{\rm eff}=s^{1/3}a$ has been used.

The suprathermal rotation number for grains spun-up by RATs is then,
\bea
\St_{\rm RAT}&=&\frac{\Omega_{\rm RAT}}{\Omega_{T}}\nonumber\\
&\simeq &1800 \hat{\rho}^{1/2}s^{5/6}a_{-5}^{7/2}\left(\frac{\bar{\lambda}}{1.2\mum}\right)^{-2}
\left(\frac{\gamma U}{n_{1}T_{2}}\right)\nonumber\\
&&\times \left(\frac{1}{1+F_{\rm IR}}\right),\label{eq:S_RAT1}
\ena
and
\bea
\St_{\rm RAT}&\simeq& 1.56\times 10^{5}\hat{\rho}^{1/2}s^{-1/6}a_{-5}^{1/2}\left(\frac{\bar{\lambda}}{1.2\mum}\right) \left(\frac{\gamma U}{n_{1}T_{2}}\right)\nonumber
\\
&&\times \left(\frac{1}{1+F_{\rm IR}}\right).\label{eq:S_RAT2}
\ena
which reveals that the suprathermal number increases rapidly with the grain size as $a^{7/2}$ for $s^{1/3}a<a_{\rm trans}$ and as $a^{1/2}$ as $s^{1/3}a>a_{\rm trans}$.

\subsection{Radiative precession around $\bk$ and $k$-RAT alignment}
RATs can induce the radiative precession of the grain around the radiation direction ($\bk$). For a grain rotating with the angular momentum $J$, the radiative precession time around $\bk$ is given by (\citealt{2014MNRAS.438..680H}):
\bea
\tau_{k}&=&\frac{2\pi}{|d\phi/dt|}=\frac{2\pi J}{\gamma u_{\rad}\lambda a_{\rm eff} ^{2}Q_{e3}},\nonumber\\
&\simeq& 179.6
\hat{\rho}^{1/2}T_{2}^{1/2}\hat{s}^{-1/6}a_{-5}^{1/2}\St\left(\frac{1.2\mum}{\gamma \bar{\lambda}\hat{Q}_{e3}U}\right)\,
\yr,\label{eq:tauk}
\ena 
where $d\phi/dt$ is the precession rate with $\phi$ the azimuthal angle, and $\hat{Q}_{e3}=Q_{e3}/10^{-2}$ with $Q_{e3}$ the third component of RATs that induces the grain precession around $\bk$ \citep{2007MNRAS.378..910L}. 

When the radiative precession is faster than the gas damping, i.e., $\tau_{k}<\tau_{\rm gas}$, the radiation direction is an axis of grain alignment. This case is known as $k$-RAT alignment \citep{2007MNRAS.378..910L}. Due to the dependence of $\tau_{k}$ on the grain angular momentum, the $k$-RAT alignment will be determined by the value of $J$, as shown in \cite{Hoang.20220o}.

\subsection{Larmor precession of HAC and the $B$-RAT alignment}
In the presence of an ambient magnetic field, the interaction of the HAC grain magnetic moment (Eq. \eqref{eq:muBar}) with the external static magnetic field causes the regular precession of $\bJ$ around $\Bv$, namely Larmor precession (see Figure \ref{fig:kB_RAT}). Therefore, the grain angular momentum experiences two precession processes of radiative and Larmor precession. When the rate of the Larmor precession is faster than the former one, the axis of grain alignment can be changed from $\kv$ to $\Bv$, which is known as $B$-RAT alignment (see e.g., \citealt{Hoang.20220o}).

The Larmor precession time, denoted by $\tau_{B}$, is given by
\bea
\tau_{B}&=&\frac{2\pi}{|d\phi/dt|}=\frac{2\pi I_{\|}\Omega}{|\mu_{\Bar}|B}=\frac{2\pi |\gamma_{n}|I_{\|}}{\chi_{n}(0)VB},\nonumber\\
&=&7.1\times 10^{3}\hat{\rho}\left(\frac{g_n}{3.1}\right)\frac{a_{-5}^{2}}{\hat{\chi}_{n}\hat{B}}~\yr,
\label{eq:tauB}
\ena
where $\hat{B}=B/5\,\mu$G and $\hat{\chi}_{n}=\chi_{n}(0)/10^{-11}$ are the normalized magnetic field and nuclear magnetic susceptibility (see Eq.~\eqref{eq:chi_n}), respectively. 

Comparing the Larmor precession to the gas rotational damping, one can see that $\tau_{B}\ll \tau_{\rm gas}$. Therefore, the Larmor precession is much faster than gas collisions, so that HAC grains can be aligned with $\Bv$. 

\subsection{A model of RAT alignment}

In addition to the spin-up, spin-down, and precession effects, RATs have an aligning torque component that acts to align the grain angular momentum with $\kv$ (or $\bv$) at a high$-J$ attractor and a low$-J$ attractor \citep{2007MNRAS.378..910L, Hoang:2008gb,2016ApJ...831..159H}. 
Therefore, a fraction of grain ensemble can be aligned at the high$-J$ attractor, denoted by $f_{\rm high-J}$, and the $1-f_{\rm high-J}$ fraction of grains are aligned at the low$-J$ attractor when the grain randomization by gas collisions is disregarded \citep{2014MNRAS.438..680H}. This process can occur on a timescale shorter than the gas damping, which is the so-called {\it fast} alignment \citep{2007MNRAS.378..910L,Lazarian.2020}.

Grains aligned at the high$-J$ attractor rotate suprathermally with $\Omega\sim \Omega_{\rm RAT}$ for RATs, corresponding to the suprathermal number of $\St_{\rm high-J}>1$. Grains at the low$-J$ attractor rotate with $\Omega\sim \Omega_{T}$ or $\St_{\rm low-J}\sim 1$. When the gas randomization effect is taken into account, grains at low$-J$ attractors are randomized and they are eventually transported to the high$-J$ attractor after a timescale greater than the gas damping time, which is usually called {\it slow} alignment \citep{Hoang:2008gb,Lazarian.2020}. Without high$-J$ attractors, grains are cycling between the low$-J$ attractor and high$-J$ repellor point and have an average low degree of alignment \citep{Hoang:2008gb,2016ApJ...831..159H,Lazarian.2020}. 
Therefore, as in \citep{Hoang.20220o}, in the following, we used this idealized model of external alignment by RATs to evaluate the efficiency of internal and external alignment for carbonaceous grains.

We note that the RAT paradigm above is valid for grains that have fast precession around $\kv$ and $\Bv$, so that one can take the averaging of RATs over the precession to solve the dynamical equations \citep{2007MNRAS.378..910L,2008MNRAS.388..117H}. This assumption is satisfied for silicate grains having fast Larmor precession due to large magnetic susceptibility. However, carbon dust has a rather low susceptibility and is expected to experience rather slow Larmor precession. Thus, we will need to evaluate whether carbonaceous grains at low$-J$ and high$-J$ attractors are fast enough to be stably aligned against gas collisions.\footnote{A detailed study of grain alignment with slow precession requires solving the full equation of motion for three parameters $J,\Theta,\phi$ that describes $\bJ$ in the lab, which will be presented elsewhere (see Figure \ref{fig:kB_RAT}).}

\subsection{Critical size for the RAT alignment}
Efficient alignment is only achieved when grains rotate suprathermally (\citealt{2016ApJ...821...91H}). Therefore, one can define the minimum size for grain alignment by RATs by setting the suprathermal condition of $\Omega_{\rm RAT}=3\Omega_{T}$, which Equation \eqref{eq:omega_RAT1} yields
\bea
a_{\rm align}^{\rm RAT}&=&0.016s^{-5/21}\hat{\rho}^{-1/7}\left(\frac{\bar{\lambda}}{1.2\mum}\right)^{4/7}\\
&&\quad\times\left(\frac{\gamma U}{n_{1}T_{2}}\right)^{-2/7}\left(1+F_{\rm IR}\right)^{2/7}\,\mum,\label{eq:aali_RAT}
\ena
which implies that even small carbonaceous grains of $a>0.016\mum$ can be aligned in the standard ISM. Toward higher-density regions, large grains can be aligned by RATs.

\section{Internal alignment}\label{sec:internal}
In this section, we study in detail the internal alignment of grains (i.e., the alignment of the grain axis of maximum inertia with its angular momentum, see Figure \ref{fig:torque-free}) by internal relaxation and derive critical sizes for internal alignment using the RAT alignment paradigm described in the previous section \citep{Hoang.20220o}.

\subsection{Nuclear relaxation}
Nuclear magnetism can induce internal relaxation as the Barnett effect for electron spins (\citealt{1999ApJ...520L..67L}). The characteristic timescale of nuclear relaxation (NR) is given by
\bea
\tau_{\rm NR}&=&\frac{\gamma_{n}^{2}I_{\|}^{3}}{VK_{n}(\omega)h^{2}(h-1)J^{2}},\nonumber\\
&\simeq&125\,\hat{\rho}^{2}a_{-5}^{7}f(\hat{s})\left(\frac{n_e}{n_n}\right) \left(\frac{J_{d}}{J}\right)^{2}
\left(\frac{g_n}{3.1}\right)^{2}\left(\frac{2.79\mu_N}{\mu_n}\right)^{2}\nonumber\\
&&\times \left[1+\left(\frac{\omega\tau_{n}}{2}\right)^{2}\right]^{2}\,\s,
\label{eq:tau_nucl}
\ena
where $f(s)=s[(1+s^{2})/2]^{2}$ and $J_d=\sqrt{I_{\|}kT_{d}/(h-1)}$ is a characteristic angular momentum estimated at the grain temperature $T_{d}$. We note that $\tau_{\rm NR}$ increases with the grain size as $a^{7}$ and angular momentum as $J^{2}$. It also increases rapidly with the rotation frequency at $\omega>2/\tau_{n}$.

Internal alignment of grains by nuclear relaxation is established when $\tau_{\rm NR}<\tau_{\rm gas}$. Following \citep{Hoang.20220o}, the maximum grain size for internal alignment induced by nuclear relaxation can be defined by $\tau_{\rm NR}=\tau_{\rm gas}$ and takes
\bea
a_{\rm max,aJ}(\rm NR) &\simeq& 6.37 h^{1/3}\St^{1/3}\left[\frac{(n_{e}/n_{n})(g_{n}/3.1)^{2}(2.79\mu_N/\mu_n)^2}{n_{1}T_{2}^{1/2}\Gamma_{\|}}\right]^{1/6}\nonumber\\
&\times&\left[\frac{1}{1+(\omega\tau_{\rm n}/2)^{2}}\right]^{1/3} \left(\frac{(h-1)T_{\rm gas}}{T_{\rm d}}\right)^{1/6}~\mum,\label{eq:a_nucl}
\ena
which increases with the suprathermal rotation as $\St^{1/3}$ for $\omega\ll 2/\tau_{n}$ but decreases as $\St^{-1/3}$ for $\omega\gg 2/\tau_{n}$, but it decreases with the gas density as $n_{\H}^{-1/6}$ \citep{Hoang.2022}. 

Equation (\ref{eq:a_nucl}) reveals that large grains aligned at low$-J$ attractors with thermal rotation of $\St=1$ can still be aligned by nuclear relaxation with $a_{\max}(NR)\sim 5.4-0.8\mum$ for $n_{\H}\sim 10-10^{6}\cm^{-3}$. For grains aligned at high$-J$ attractors, one has $\St=\St_{\rm RAT}$. Due to its dependence on the grain size, can obtain $a_{\rm max,aJ}$ by numerically solving the equation $\tau_{\rm NR}(a)=\tau_{\rm gas}$.

\subsection{Inelastic relaxation}
Inelastic relaxation is suggested by \cite{1979ApJ...231..404P} as a mechanism to cause internal alignment. This mechanism is determined by the inelasticity of grain material and does not depend on grain magnetism. Thus, it can work for any form of carbonaceous dust, including graphite and HAC. \cite{1999MNRAS.303..673L} revisited the inelastic relaxation for a squared prism by taking into account the effect of double oscillation frequency. Later, \cite{Molina.2003} revisited the treatment of LE99 for an oblate spheroid. The characteristic relaxation can be written as (see \citealt{Hoang.20220o})
\bea
\tau_{\rm iER}\approx \frac{\mu Q}{\rho a^{2}\Omega_{0}^{3}}g(s),\label{eq:ti_LE}
\ena
where the $g$ factor is a geometrical function depending on the axial ratio as
\bea
g(s)=\frac{ 2^{3/2}7}{8}\frac{(1+s^{2})^{4}}{s^{4}+1/(1+\sigma)},\label{eq:gs}
\ena
which corresponds to $g(s)=7.0$ and $4.6$ for $s=1/2$ and $1/3$, respectively. Note that we have assumed the Poisson ratio $\sigma=0.25$ as in \cite{Molina.2003}.

Using the definition of the suprathermal number $\St=\Omega_{0}/\Omega_{T}$, Equation (\ref{eq:ti_LE}) becomes
\bea
\tau_{\rm iER}\approx \frac{\rho^{1/2}a^{11/2}\mu Q }{(kT_{\rm gas})^{3/2}\St^{3}}g_{1}(s),\label{eq:tine_LE}
\ena
where $g_{1}\approx 2.2s^{3/2}g(s)$. Using the typical parameters with $s=1/2$, one obtains
\bea
\tau_{\rm iER}\simeq 0.51\,T_{2}^{-3/2}\hat{\rho}^{1/2}a_{-5}^{11/2}\mu_{10}Q_{3}(\St)^{-3}g_1(s)~\yr\nonumber\\
\ena 
where $\mu_{10}=\mu/(10^{10}\erg/\cm^{3})$ with $\mu$ the modulus of rigidity, and $Q_{3}=Q/10^{3}$ with elastic $Q$ parameter, assuming the brick grain shape of axial ratio $s=1/2$. The typical value $Q$ is about 100 for silicate rocks and a lower value is expected for carbon dust (see \citealt{Efroimsky:2000p5384}). Here, we assume $\mu_{10}=1$ and $Q_{3}=1$ for numerical estimates.

Following \cite{Hoang.20220o}, one can obtain the critical size for internal alignment by inelastic relaxation by setting $\tau_{\rm iER} =\tau_{\rm gas}$, which reads
\bea
a_{\rm max,aJ}(\rm iER)&\simeq&1.44\,s^{2/9}\hat{\rho}^{1/9}n_{1}^{-2/9}T_{2}^{1/9}\nonumber\\
&&\,\times \mu_{10}^{-2/9}Q_3^{-2/9}\St^{ 2/3}~\mum.
\ena

For grains aligned at low$-J$ attractors by RATs with $\St\sim 1$, one obtains
\bea
a_{\rm max,aJ}^{\rm low-J}(\rm iER)&\simeq&1.44\,s^{2/9}\hat{\rho}^{1/9}n_{1}^{-2/9}T_{2}^{1/9}\nonumber\\
&&\,\times \mu_{10}^{-2/9}Q_3^{-2/9}~\mum,\label{eq:amax_aJ_lowJ}
\ena
which implies $a_{\rm max,aJ}^{\rm low-J}({\rm iER})\sim 1.44, 0.31, 0.11\mum$ for $n_{\rm H}=10, 10^{4}, 10^{6}\cm^{-3}$.


For grains aligned at high$-J$ attractors by RATs, which have suprathermal rotation $\St=\St_{\rm RAT}$, the minimum size of internal alignment is
\bea
a_{\rm min,aJ}^{\rm high-J}(\rm iER)&=&0.0026s^{-7/12}\left(\frac{\mu_{10}Q_{3}g_{1}n_1^4T_2^2}{\hat{\rho}^2}\right)^{1/6}\nonumber\\
&&\times\left(\frac{\bar{\lambda}}{1.2\,\mum}\right)(\gamma U)^{-1/2}\,\mum,\label{eq:amin_aJ_RAT}
\ena
where $a_{\rm min}$ is the minimum grain size for internal alignment which is the solution of the equation $\tau_{\rm iER}=\tau_{\gas}$. 

Similarly, the maximum size of internal alignment by inelastic relaxation is
\bea
a_{\rm max,aJ}^{\rm high-J}({\rm iER})&=&1.25\times10^{4}s^{-1/6}\left(\frac{\mu_{10}Q_{3}g_{1}n_1^4T_2^2}{\hat{\rho}^2}\right)^{-1/3}\nonumber\\
&&\times\left(\frac{\bar{\lambda}}{1.2\,\mum}\right)\gamma U\,\mum.\label{eq:amax_aJ_RAT}
\ena

Equations (\ref{eq:amin_aJ_RAT}) and (\ref{eq:amax_aJ_RAT}) imply $a_{\rm min,aJ}^{\rm high-J}({\rm iER})=0.026, ...$  and $a_{\rm max,aJ}^{\rm high-J}({\rm iER})\sim 10^4\mum$ for $U=1, n_{\H}\sim 10^{3}\cm^{-3}$ and $T_{\rm gas}=10$ K. For the diffuse ISM, carbonaceous grains have small sizes of $a<1\mum$. Therefore, the entire population of carbon dust can have efficient internal alignment by fast inelastic relaxation.

\begin{figure}
\includegraphics[width=0.5\textwidth]{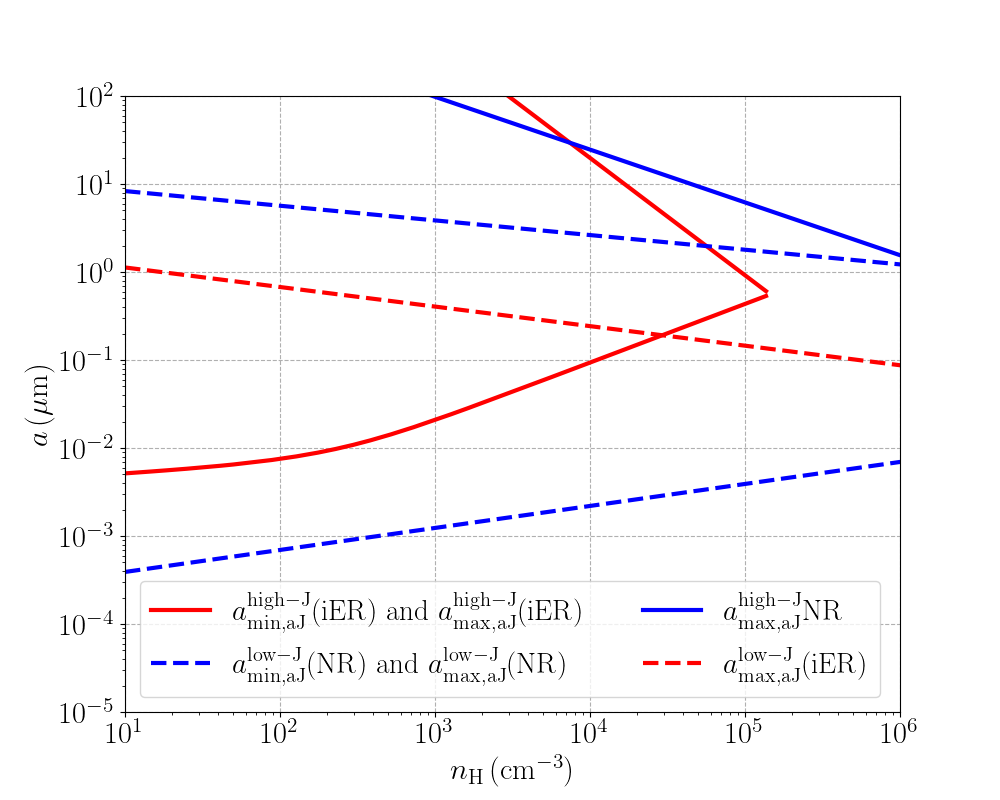}
\caption{Grain size for efficient internal alignment from inelastic relaxation ($a^{\rm high-J}_{\rm min,aJ}(\rm iER)$ and $a^{\rm high-J}_{\rm max,aJ}(\rm iER)$, solid red lines) and from nuclear relaxation ($a_{\rm max,aJ}^{\rm high-J}(\rm NR)$, solid blue line) versus the gas density, assuming grains rotating suprathermally at high-J attractors by RATs with $U=1, \gamma=0.3$, and $\bar{\lambda}=1.2\mum$. 
Grain sizes for efficient internal alignment due to inelastic relaxation ($a^{\rm low-J}_{\rm max,aJ}({\rm iER})$, dashed red line) and nuclear relaxation ($a^{\rm low-J}_{\rm max,aJ}({\rm NR})$, dashed blue line) for grains aligned at low$-J$ attractors are also shown.}
\label{fig:fg_aRAT}
\end{figure}

Figure \ref{fig:fg_aRAT} shows in detail the range of grain sizes that have internal alignment by nuclear relaxation and inelastic relaxation for the different gas densities, assuming grains aligned at high-J (solid lines) and low-J attractors (dashed lines). As shown, the range of grains with internal alignment is broader for grains rotating suprathermally at high-J attractors. In the considered range of gas density, nuclear relaxation can induce internal alignment for a wide range of grain sizes (see blue solid and dashed lines). Inelastic relaxation is only efficient for $n_{\H}<10^{5}\cm^{-3}$ if grains align at high-J attractors (see solid red lines) and much less efficient for grains at low-J attractors (see dashed red line).

\begin{figure*}
\includegraphics[width=0.5\textwidth]{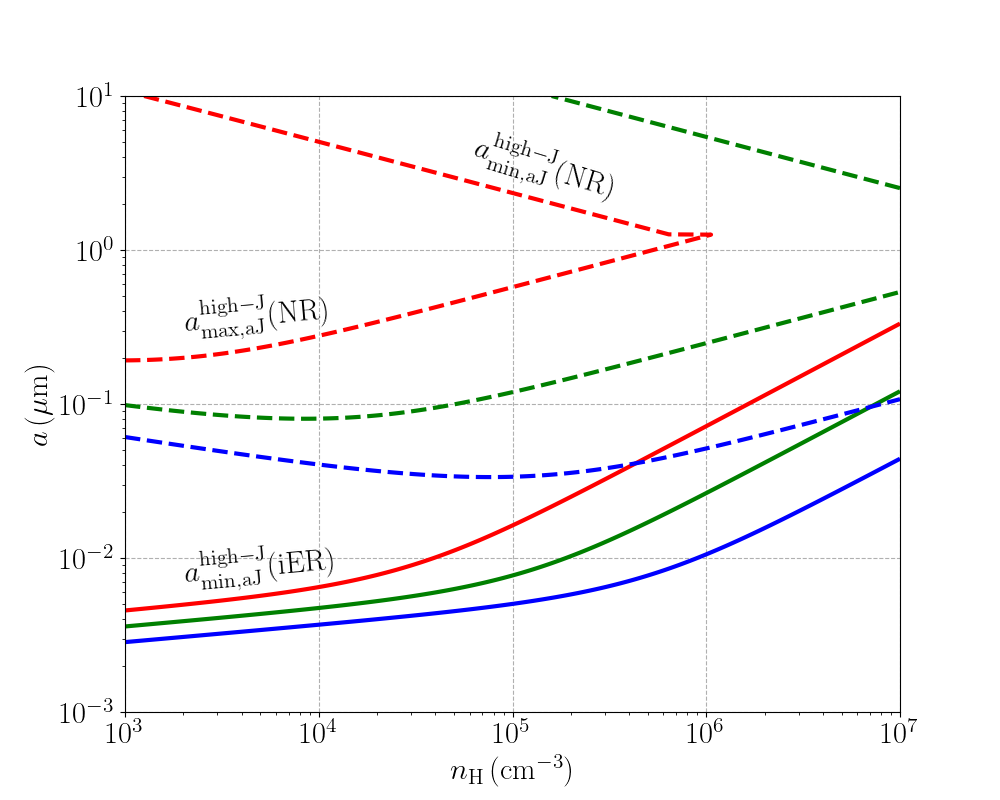}
\includegraphics[width=0.5\textwidth]{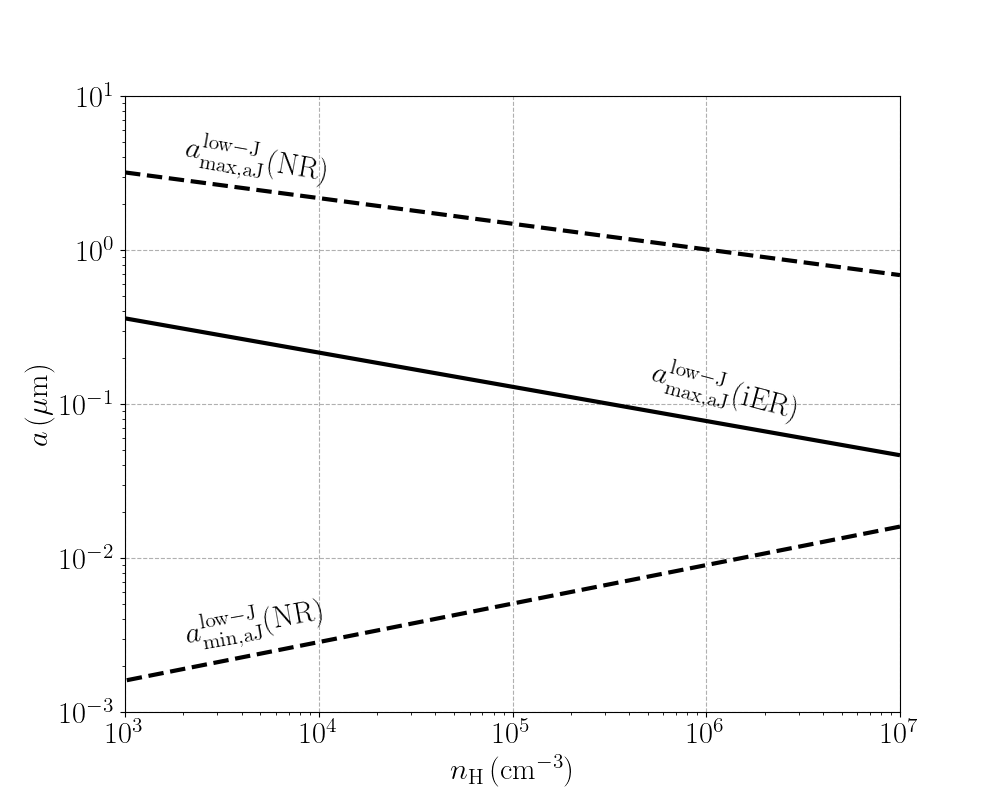}
\caption{Left panel: Grain size for efficient internal alignment from inelastic relaxation ($a^{\rm high-J}_{\rm min,aJ}(\rm iER)$, solid lines) and from nuclear relaxation ($a^{\rm high-J}_{\rm min,aJ}(\rm NR)$ and $a^{\rm high-J}_{\rm max,aJ}(\rm NR)$, dashed lines) versus the gas density assuming grains aligned at high-J attractors by RATs with $U=10^{3}$, $10^4$, and $10^5$ ($\gamma=1$ and $\bar{\lambda}=2.5$ $\mu$m). Right panel: Same as the left panel but for grains aligned at low-J attractors with $\Omega=\Omega_T$.}
\label{fig:amax_aJ_RAT}
\end{figure*}

Figure \ref{fig:amax_aJ_RAT} (left panel) shows critical sizes for grains aligned at high$-J$ attractors for grains subject to strong radiation fields of $U=10^{3}-10^{5}$. For nuclear relaxation, we note that the notations might be quite confusing in this case as $a^{\rm high-J}_{\rm max,aJ}(\rm NR)\leq a^{\rm high-J}_{\rm min,aJ}(\rm NR)$. This is because efficient internal alignment is achieved for $a\leq a^{\rm high-J}_{\rm max,aJ}(\rm NR)$ and $a\geq a^{\rm high-J}_{\rm min,aJ}(\rm NR)$. In other words, there is a gap in grain size ($ a^{\rm high-J}_{\rm max,aJ}(\rm NR) < a < a^{\rm high-J}_{\rm min,aJ}(\rm NR)$) where grains might not have internal alignment via nuclear relaxation. This gap gets wider with increasing the radiation field strength (less efficient internal alignment via nuclear relaxation for grains with typical sizes) due to the suppression of the nuclear magnetic susceptibility with fast rotation (dashed lines). In contrast, inelastic relaxation becomes more efficient with increasing radiation field strength due to faster rotation. We could see that $a^{\rm high-J}_{\rm min,aJ}({\rm iER}$ (solid lines) gets smaller for larger values of $U$. In general, we also see that $a^{\rm high-J}_{\rm max,aJ}({\rm NR}) \gtrsim a^{\rm high-J}_{\rm min,aJ}({\rm iER})$ over a large range of gas density given $U\gtrsim 10^3$-$10^5$. This means that inelastic relaxation is more important for internal alignment of high$-J$ carbonaceous grains than nuclear relaxation in strong radiation fields.

Figure \ref{fig:amax_aJ_RAT} (right panel) shows the critical sizes for grains aligned at low-$J$ attractors. Grains with sizes larger than $a_{\rm max,aJ}({\rm iER})$ (solid line) cannot have efficient internal alignment by inelastic relaxation. We show also that internal relaxation is only efficient within a certain range of grain sizes ($a^{\rm high-J}_{\rm min,aJ}(\rm NR) < a < a^{\rm high-J}_{\rm max,aJ}(\rm NR)$, dashed lines). For low$-J$ carbonaceous grains, both nuclear relaxation and inelastic relaxation are important for internal alignment.



\section{External Alignment}\label{sec:external}
Here we study in detail the external alignment of grains (i.e., the alignment of grain angular momentum with a preferred axis) in the space, including the radiation direction ($\kv$), magnetic field ($\Bv$), and electric field (${\bf E}$). We will derive critical sizes for the external alignment of carbonaceous grains using the RAT alignment paradigm with low$-J$ and high$-J$ attractors \citep{Hoang.20220o}.

\subsection{The minimum size of RAT alignment}
In the RAT paradigm, the minimum size of efficient grain alignment by RATs is defined by the suprathermal rotation conditions of $\Omega_{\rm RAT}=3\Omega_{T}$, which is given by (see \citealt{Hoang:2021ct,Hoang.20220o}):
\bea
a_{\rm align}^{\rm RAT}&\simeq&0.016s^{-5/21}\hat{\rho}^{-1/7}\left(\frac{\bar{\lambda}}{1.2\mum}\right)^{4/7}\left(\frac{\gamma U}{n_{1}T_{2}}\right)^{-2/7}\nonumber\\
&\times&\left(\frac{1}{1+F_{\rm IR}}\right)^{-2/7} \mum,\label{eq:aali_RAT}
\ena
which implies $a_{\rm align}\approx 0.04\mum$ for the typical ISM conditions of $n_{\H}=30\cm^{-3}, T_{\rm gas}=100\K, \gamma=0.1, U=1$ with $F_{\rm IR}< 1$ from Equation (\ref{eq:FIR}).\footnote{There is a misprint in the last term of Eq. 65 in \cite{Hoang.20220o} where the slope $2/7$ should read $-2/7$.} Grains smaller than $a_{\rm align}$ rotate thermally and have negligible alignment efficiency (\citealt{2016ApJ...831..159H,2016ApJ...821...91H}).

\subsection{The $k$-RAT alignment of graphite}
In the absence of magnetic fields, dust grains subject to an anisotropic radiation field experience RATs and collisions with gas species. If the radiative precession of grains around $\kv$ induced by RATs is faster than the gas damping, then, grains can be stably aligned along the radiation direction (aka. $k$-RAT mechanism). As shown in Equation (\ref{eq:tauk}), the timescale of radiative precession depends on the grain angular momentum. Thus, the timescale of radiative precession for grains aligned at high$-J$ and low$-J$ attractors is significantly different. Following \cite{Hoang.20220o}, the radiative precession timescale for grains aligned at low$-J$ attractors with $\St=1$ is
\bea
\tau_{k}^{\rm low-J}=179.6\hat{\rho}^{1/2}T_{2}^{1/2}\hat{s}^{-1/6}a_{-5}^{1/2}\left(\frac{1.2\mum}{\gamma\bar{\lambda}\hat{Q}_{e3}U}\right)
\yr,\label{eq:tauk_lowJ}
\ena 
and for grains aligned at high$-J$ with $\St=\St_{\rm RAT}$, the radiative precession time becomes
\bea
\tau_{k}^{\rm high-J}&=&2.89\times 10^{5}s^{2/3}\left(\frac{\hat{\rho}a_{-5}^{4}}{\hat{Q}_{e3}}\right)\left(\frac{1}{n_{1}T_{2}^{1/2}}\right)\left(\frac{\bar{\lambda}}{1.2\,\mum}\right)^{-3}\nonumber\\
&&\times \left(\frac{1}{1+F_{\rm IR}}\right)\yr,\label{eq:tauk_highJ1}
\ena
for $a<a_{\rm trans}$, and 
\bea
\tau_{k}^{\rm high-J}&=&2.49\times 10^{7}s^{-1/3}\left(\frac{\hat{\rho}a_{-5}}{\hat{Q}_{e3}}\right)\left(\frac{1}{n_{1}T_{2}^{1/2}}\right)\nonumber\\
&&\times\left(\frac{1}{1+F_{\rm IR}}\right)\yr,\label{eq:tauk_highJ2}
\ena
for $a>a_{\rm trans}$.

Comparing Equations (\ref{eq:tgas}) and (\ref{eq:tauk_lowJ}), one can see that $\tau_{k}^{\rm low-J}\ll \tau_{\rm gas}$ for grains at low$-J$ attractors. Therefore, $\kv$ can be the axis of grain alignment at low$-J$ attractors. Similarly, for grains aligned at high$-J$ attractors, comparing Equations (\ref{eq:tgas}) and (\ref{eq:tauk_highJ1}) reveals that $\tau_{k}^{\rm high-J}\gg \tau_{\rm gas}$ for $a\gtrsim 10^{-1}\,\mum$, assuming the typical parameters of the interstellar radiation field of $U=1$, $\gamma=0.3$ and $\bar{\lambda}=1.2\,\mum$. Therefore, $\kv$ cannot be the stable axis of alignment for sufficiently large grains at high-J attractors. In the other words, the alignment of large graphite grains via $k$-RAT can only occur at low$-J$ attractors because the high$-J$ attractor is not stable due to gas collisions.

\begin{figure*}
\includegraphics[width=0.5\textwidth]{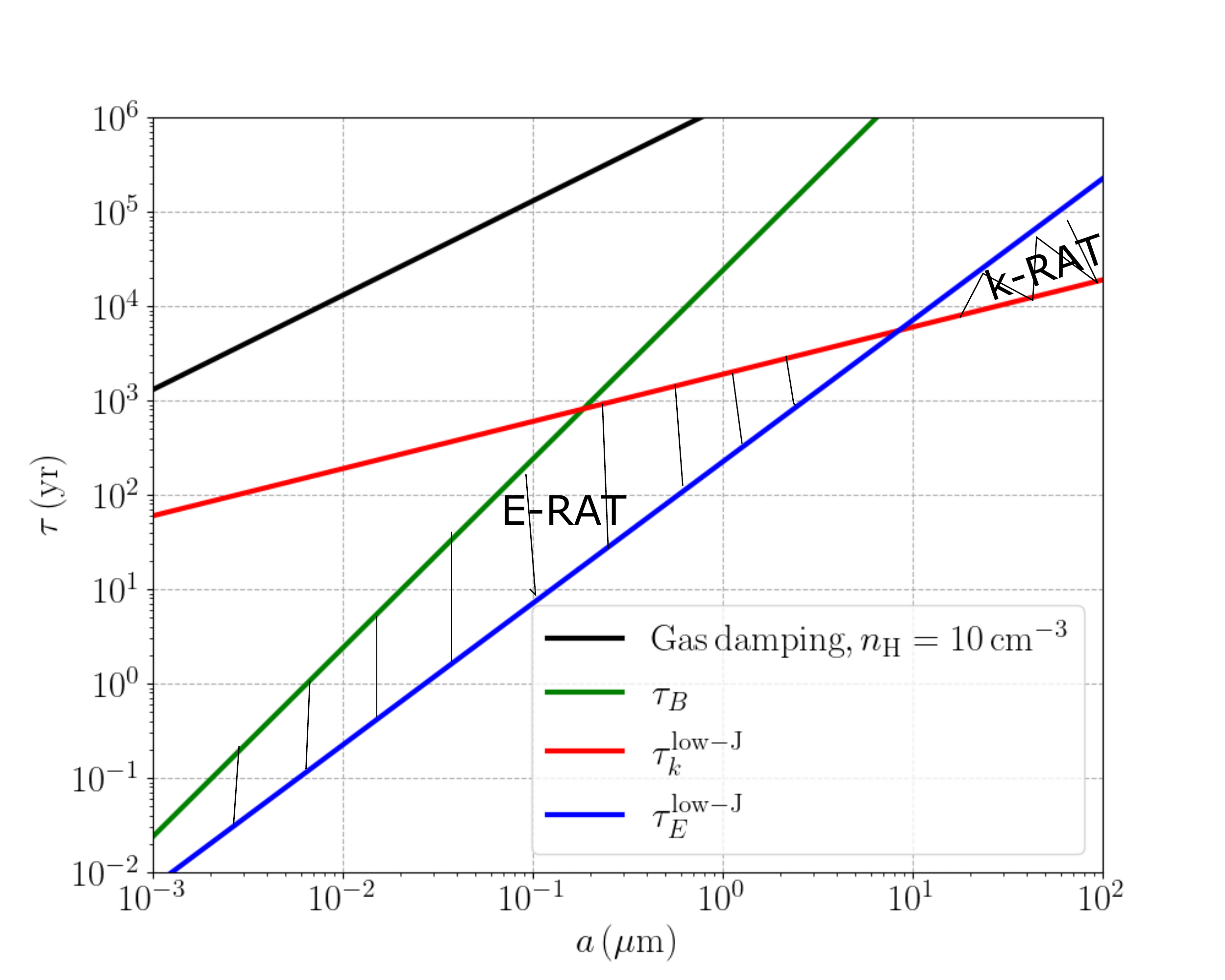}
\includegraphics[width=0.5\textwidth]{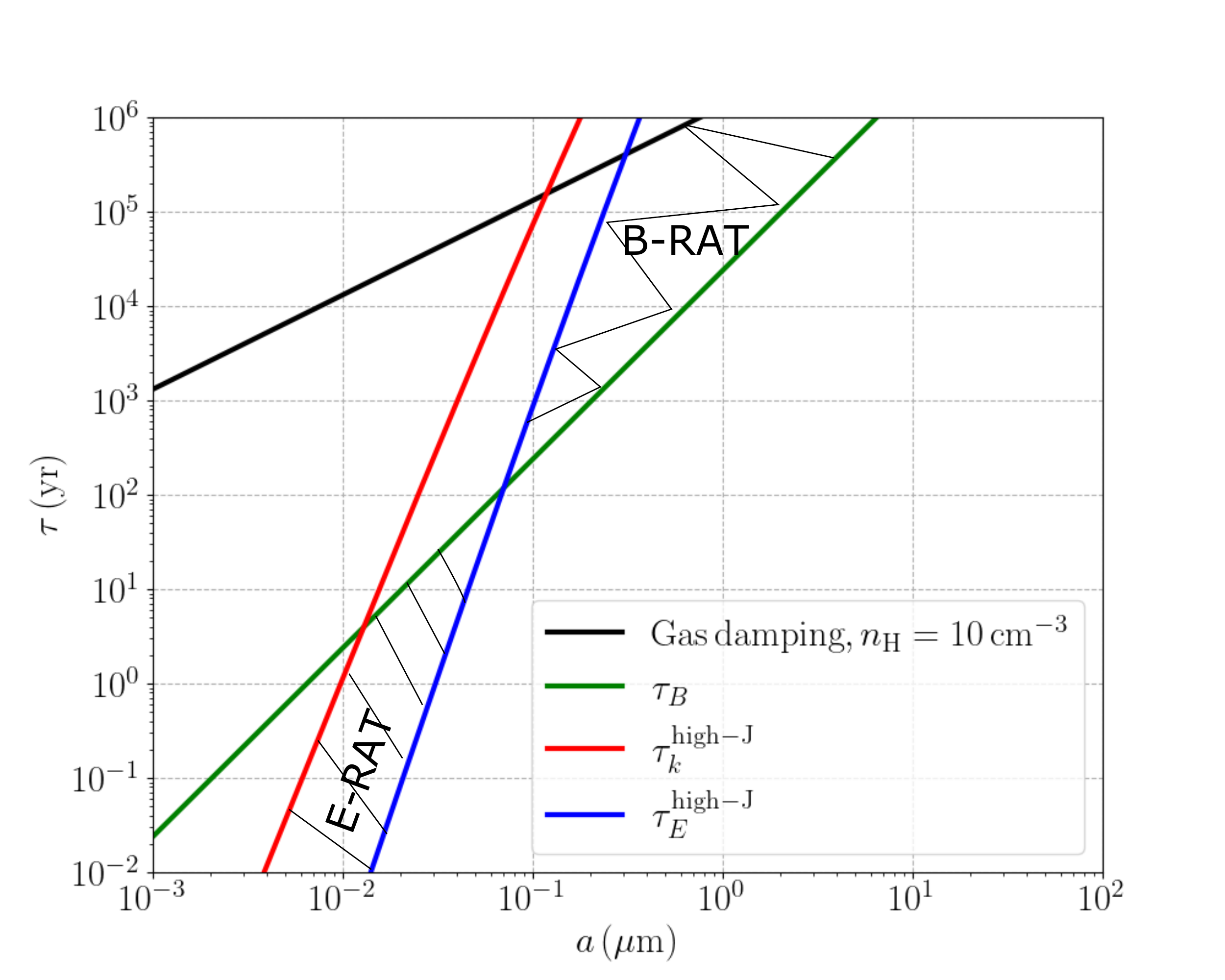}
\caption{Left panel: Timescales for external alignment via $k$-RAT (red lines), $B$-RAT (green lines), and $E$-RAT (blue lines) versus grain size. Results are shown for grains at both low$-J$ (left panel), assuming typical values of magnetic field strength $B=5\,\mu$G, gas density $n_{\rm H}=10$ cm$^{-3}$ in the diffuse ISM. The gas damping timescale is also overlaid for comparison (black lines). For $a<10\mum$, the electric precession is faster (blue line) and the grains align via the $E$-RAT mechanism. Right panel: Same as the left one but for grains aligned at high-J attractors, assuming $U=1$, $\gamma=0.3$ and $\bar{\lambda}=1.2\,\mum$. The Larmor precession is faster for $a>0.05\mum$ so that grains align via $B$-RAT mechanism. }
\label{fig:tau_kBE}
\end{figure*}

\subsection{The $k$-RAT vs $B$-RAT alignment of HAC}
Unlike diamagnetic graphite, HAC  grains with nuclear paramagnetism experience simultaneous radiative and Larmor precession. The axis of grain alignment can be changed from $\kv$ ($k$-RAT) to $\Bv$ ($B$-RAT) if the Larmor precession is faster than the radiative precession.

From Equations \eqref{eq:tauB}, \eqref{eq:tauk_lowJ}, \eqref{eq:tauk_highJ1}, and \eqref{eq:tauE_highJ2}, one can see that $\tau_{B}\gtrsim \tau_{k}^{\rm low-J}$ for grains of size $a\gtrsim 0.1\,\mum$ at low$-J$ attractors, but $\tau_{B}\lesssim \tau_{k}^{\rm high-J}$ for grains of size $a\gtrsim 0.01\,\mum$ at high$-J$ attractors (see also Figure \ref{fig:tau_kBE}). Therefore, sufficiently large grains at low$-J$ attractors can have $k$-RAT, but HAC can have $B$-RAT at high$-J$ attractors.

One can also derive the minimum size for grain alignment via $k$-RAT (or the maximum size for $B$-RAT) alignment at low$-J$ attractor using the condition $\tau_{k}^{\rm low-J,high-J}=\tau_{B}$, which yields
\bea
a^{\rm low-J}_{\rm min,Jk}&\equiv& a_{\rm max,JB}^{
\rm low-J}=3.7\times 10^{-4}s^{-1/9}\nonumber\\
&\times&\left(\frac{\hat{\chi}_n\hat{B}T_2^{1/2}}{\gamma U\rho^{1/2}Q_{e3}}\right)^{2/3}\left(\frac{\bar{\lambda}}{1.2\,\mum}\right)^{-2/3}\,\mum
\ena
for grains at low$-J$ attractors and
\bea
a^{\rm high-J}_{\rm min,Jk} &\simeq&a_{\rm max,JB}^{
\rm high-J}=3.51s^{-1/3}\left(\frac{n_1T_2^{1/2}Q_{e3}}{\hat{\chi}_n\hat{B}}\right)^{-1}\nonumber\\
&&\times \left(1+F_{\rm IR}\right)\,\mum
\ena
for grains at high-J attractors where $\tau_{k}^{\rm high-J}$ from Equation (\ref{eq:tauk_highJ2}) has been used. 

\begin{figure}
\includegraphics[width=0.5\textwidth]{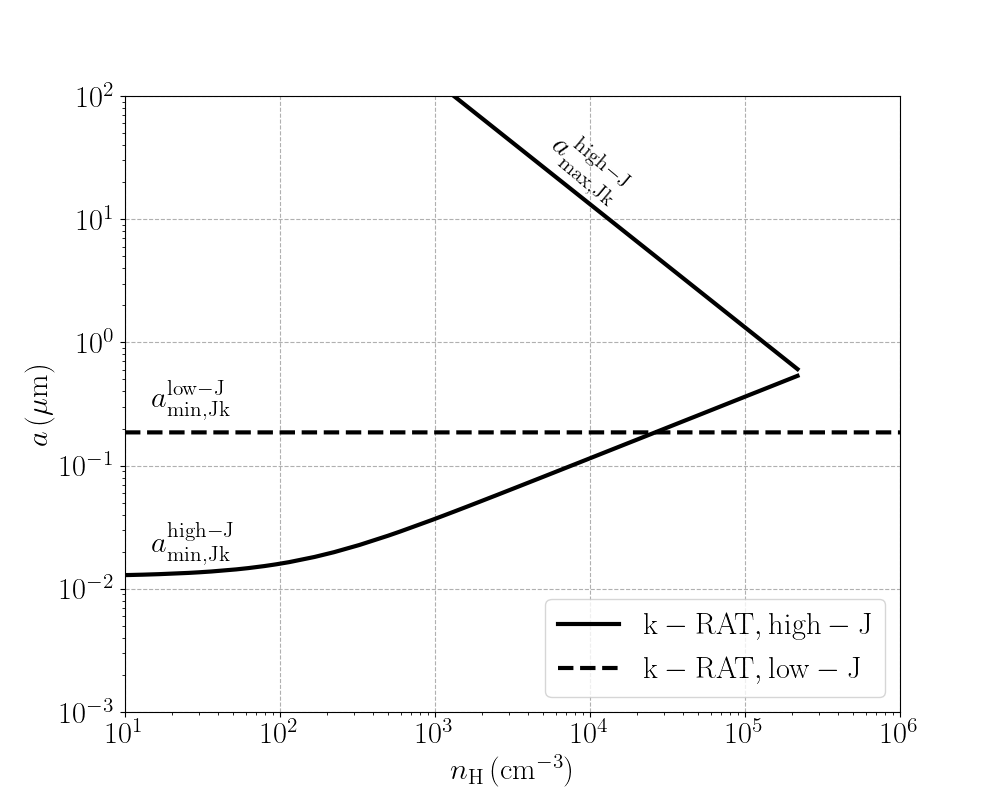}
\caption{Grain size for efficient external alignment via $k$-RAT at both low$-J$ ($a^{\rm low-J}_{\rm min,Jk}$, dashed line) and high$-J$ ($a^{\rm high-J}_{\rm min,Jk}$ and $a^{\rm high-J}_{\rm max,Jk}$, solid line) attractors. Grains at high-$J$ attractor are assumed to rotate suprathermally via RAT in the ISRF with $U=1$, $\gamma=0.3$, and $\bar{\lambda}=1.2\,\mum$.}
\label{fig:fg_a_Jk_B=5_ISM}
\end{figure}

Figure \ref{fig:fg_a_Jk_B=5_ISM} shows the exact values of these critical sizes obtained by solving numerically the equation $\tau_k=\tau_{\rm gas}$. Grain size for efficient external alignment via $k$-RAT at both low$-J$ ($a^{\rm low-J}_{\rm min,Jk}$, dashed line) and high$-J$ ($a^{\rm high-J}_{\rm min,Jk}$ and $a^{\rm high-J}_{\rm max,Jk}$, solid line) attractors are presented. Note that grains at high-$J$ attractor are assumed to rotate suprathermally via RATs in the ISRF with $U=1$, $\gamma=0.3$, and $\bar{\lambda}=1.2\,\mum$. For grains at low-J attractors, large grains of $a>0.2\mum$ can be aligned via $k$-RAT (dashed line), but smaller grains are aligned via $B$-RAT.

We now check whether the $B$-RAT alignment can be stable against gas collisions by estimating the ratio of the Larmor precession time to the gas damping time
\bea
\frac{\tau_{B}}{\tau_{\rm gas}}=2.7\times 10^{-2}s^{-1}a_{-5}\left(\frac{n_{1}T_{2}^{1/2}\Gamma_{\|}}{\hat{\chi}_{n}\hat{B}}\right),
\ena
which yields the magnetic alignment against gas collisions when $\tau_{B}/\tau_{\gas}<1$ or
\bea
a_{max,JB}^{\rm coll}<1.9\hat{s}\left(\frac{\hat{\chi}_{n}\hat{B}}{n_{1}T_{2}^{1/2}\Gamma_{\|}}\right)\mum.\label{eq:amax_JB_gas}
\ena

Thus, for typical parameters of dust grains, HAC grains of size $a<1.9\mum$ in the diffuse ISM can still be stably aligned via $B$-RAT. Note that this might not remain true if the zero-frequency susceptibility of proton nuclei $\chi_n(0)\ll 10^{-11}$, which corresponds to the case where either interstellar carbonaceous grains are purely in the form of graphite or they have only a thin layer of HAC ($n_n\ll 10^{22}$ cm$^{-3}$, see Eq.~\eqref{eq:chi_n}).


Table \ref{tab:align} summarizes our results on internal alignment and external alignment of carbonaceous grains of size $a\gtrsim10^{-5}\,\mum$ in the diffuse ISM. Internal alignment can be efficient for both nuclear and inelastic relaxation. However, external alignment cannot be stable at high$-J$ attractors, and only grains align at low$-J$ attractors.

\begin{center}
\begin{table*}
\caption{Alignment of carbon dust of size $a\gtrsim0.1\,\mum$ in the diffuse ISM assuming zero-frequency susceptibility of proton nuclei $\chi_{n}(0)\simeq 3\times 10^{-10}$ and the ISRF with $U=1$, $\gamma=0.3$, and $\bar{\lambda}=1.2\,\mum$.}\label{tab:align}
\begin{tabular}{|c|cc|cc|cc|}
\hline
{\it Internal Alignment} &  \multicolumn{3}{b{3.65cm}|}{\it\centering Inelastic Relaxation} & \multicolumn{3}{b{3.65cm}|}{\it\centering Nuclear Relaxation} \cr
\hline
{\it High$-J$}  & \multicolumn{3}{c|}{Fast} & \multicolumn{3}{c|}{Fast} \cr
\hline
{\it Low$-J$}  & \multicolumn{3}{c|}{Fast} & \multicolumn{3}{c|}{Fast} \cr
\hline
{\it External Alignment} & \multicolumn{2}{p{2cm}|}{\it\centering Radiative Precession} &  \multicolumn{2}{p{2cm}|}{\it\centering Larmor Precession} &  \multicolumn{2}{p{2cm}|}{\it\centering Electric Precession} \cr
\hline
{\it High$-J$}  & \multicolumn{2}{c|}{$\tau_{k}>\tau_{\gas}$ (Not Align)} & \multicolumn{2}{c|}{$\tau_{B}<\tau_{k}$ ($B$-RAT)}  & \multicolumn{2}{c|}{$\tau_{E}<\tau_{\rm gas}$($E$-RAT)} \cr
\hline
{\it Low$-J$}  & \multicolumn{2}{c|}{$\tau_{k}<\tau_{\gas}$ ($k$-RAT)} & \multicolumn{2}{c|}{$\tau_{B}\gtrsim\tau_{k}$($k$-RAT)} & \multicolumn{2}{c|}{$\tau_{E}<\tau_{\gas}$ $E$-RAT)} \cr
\hline
\end{tabular}
\end{table*}
\end{center}

\subsection{The $k$-RAT vs. $E$-RAT alignment}
Previous studies \citep{2014MNRAS.438..680H,Lazarian.2020bwy} suggest that carbonaceous grains may be aligned along the electric field (${\bf E}$) due to the fast precession of the grain dipole moment with respect to the electric field. Here, we quantify the alignment along the electric field within the RAT paradigm with grains rotating on two possible states of low-J and high-J attractors.

A dust grain drifting through the magnetized interstellar gas of velocity $v_{\rm d}$ experiences an induced electric field, 
\bea
E_{\rm ind}&=&\frac{1}{c}v_{d}B\nonumber\\
&\simeq &3.3\times 10^{-12}\left(\frac{v_{d}}{0.1~\rm km\s^{-1}}\right)\left(\frac{B}{10\mu  G}\right) ~{\rm stat~V}\cm^{-1},~~~
\label{eq:Eind}
\ena
where $v_{d}$ is normalized to the typical value of grain velocity in the diffuse ISM achieved by acceleration in gyroresonance MHD turbulence \citep{Hoang:2012cx}.

The interaction between the grain dipole moment (e.g., due to asymmetric charge distribution) with the induced electric field induces an electric torque of $\bp\times {\bE}$, which causes the grain to experience electric precession around ${\bE}$ with a period of (see \citealt{2014MNRAS.438..680H})
\bea
\tau_{E}={\frac{2\pi}{|d\phi/dt|}}=\frac{ 2\pi I_{\|}\Omega}{p_{Z}E},\label{eq:tauE1}
\ena
where the electric dipole moment is taken along $\bJ$. In a general case, the electric dipole moment makes an angle with the grain axis of maximum inertia $\ahat_{1}$. However, due to the fast spinning around $\ahat_{1}$, the net dipole moment averaged over the fast spinning is only directed along $\ahat_{1}$, which is along $\bJ$ due to efficient inelastic relaxation.

Using the dipole moment from Equation (\ref{eq:pZ}) and induced electric field from Equation \eqref{eq:Eind} for Equation (\ref{eq:tauE1}), one obtains
\bea
\tau_{E}&=&10 \hat{\rho}^{1/2}T_2^{1/2}s^{1/2}a_{-5}^{3/2}\St  \left(\frac{\epsilon}{0.01}\right)^{-1}\nonumber\\
&\times&\hat{B}^{-1}\left(\frac{\phi}{0.3\, \rm V}\right)^{-1}\left(\frac{v_{d}}{0.1\km \s^{-1}}\right)^{-1}\yr. 
\label{eq:tauE}
\ena
For grains rotating suprathermally at high$-J$ attractors by RATs, $\St=\St_{\rm RAT}$, the above equation becomes
\bea
\tau_{E}^{\rm high-J}&=&1.8\times10^4\hat{\rho}s^{4/3}a_{-5}^{5}\left(\frac{\epsilon}{0.01}\right)^{-1}
\left(\frac{\phi}{0.3\, \rm V}\right)^{-1}\nonumber\\
&\times& \hat{B}^{-1}\left(\frac{v_{d}}{0.1\,\rm km\s^{-1}}\right)^{-1}\nonumber\\
&\times&\left(\frac{\bar{\lambda}}{1.2\mum}\right)^{-2}
\left(\frac{\gamma U}{n_{1}T_{2}^{1/2}}\right)\left(\frac{1}{1+F_{\rm IR}}\right)
\yr, 
\label{eq:tauE_highJ1}
\ena
for $a<a_{\rm trans}$, and
\bea
\tau_{E}^{\rm high-J}&=&1.56\times 10^{6}\hat{\rho}s^{1/3}a_{-5}^2\left(\frac{\epsilon}{0.01}\right)^{-1}
\left(\frac{\phi}{0.3\, \rm V}\right)^{-1}\nonumber\\
&\times&\hat{B}^{-1} \left(\frac{v_{d}}{0.1\,\rm km\s^{-1}}\right)^{-1}\nonumber\\
&\times&
\left(\frac{\bar{\lambda}}{1.2\mum}\right) \left(\frac{\gamma U}{n_{1}T_{2}^{1/2}}\right)\left(\frac{1}{1+F_{\rm IR}}\right)\yr,
\label{eq:tauE_highJ2}
\ena
for large grains of $a>a_{\rm trans}$. 

Comparing $\tau_{E}$ to $\tau_{\rm gas}$ (Eq. \eqref{eq:tgas}), one can see that grains at high$-J$ attractors have $\tau_{E}\gg \tau_{\gas}$ for large grains of $a>a_{\rm trans}$, but $\tau_{E}\ll \tau_{\gas}$ for $a<a_{\rm trans}$. Moreover, for grains at low$-J$ attractors, $\St\sim 1$, then, $\tau_{E}\gg \tau_{\gas}$. Therefore, grains cannot be efficiently aligned at high$-J$ attractors, but they can be weakly aligned with ${\bf E}$ at low$-J$ attractors.

Figure \ref{fig:tau_kBE} depicts the characteristic timescales of the grain precession around $\kv, \bE, \Bv$ versus the grain size with the gas damping timescale overlaid. We present results for grains aligned at both low$-J$ (left panel) and high$-J$ (right panel) attractors, assuming the drift velocity $v_{d}=0.1\km\s^{-1}$, the magnetic field strength $B=5\,\mu$G, and gas density $n_{\rm H}=10$ cm$^{-3}$ for the diffuse ISM. In the case of low-J attractors, the electric precession is faster for $a<10\mum$ (blue line) and these grains can align via the $E$-RAT mechanism. However, for the high-J attractor case, Larmor precession is fastest for $a>0.05\mum$, and these large grains can align via the $B$-RAT mechanism. The $E$-RAT alignment is possible for grains between $0.01-0.05\mum$ (see green line).

We now determine the critical size for grains aligned with $E$-RAT vs $k$-RAT by comparing $\tau_{E}$ (Eq. \ref{eq:tauE}) with $\tau_{k}$ (\ref{eq:tauk}). One obtains the maximum size for $E$-RAT alignment,
\bea
a_{\rm max,JE}^{\rm low-J,high-J}
&\simeq& 1.6s^{-2/3}\hat{B}^{-1}\left(\frac{\epsilon}{0.01}\right)
\left(\frac{\phi}{0.3 \rm V}\right)\nonumber\\
&\times&\left(\frac{v_{d}}{0.1\,\rm km\s^{-1}}\right)\left(\frac{1.2\,\mum}{\gamma U\bar{\lambda}\hat{Q}_{e3}}\right)\mum~~~.\label{eq:max_JE}
\ena
for grains at both low$-J$ and high$-J$ attractors. Note that this is applicable for both low and high$-J$ attractors because both $\tau_E$ and $\tau_k$ are proportional to ${\rm St}$.

The maximum grain size for $E$-RAT is larger for faster grain drift through the gas ($v_{d}$).

\subsection{Effect of Nuclear Magnetism on Paramagnetic Relaxation}
\cite{1951ApJ...114..206D} (D-G) first suggested the paramagnetic relaxation of interstellar grains as an alignment mechanism to explain starlight polarization. The basic idea of this D-G mechanism is as following. Consider a grain of paramagnetic material, with susceptibility $\chi(0)$, rotating with $\bJ$ not aligned with the ambient magnetic field $\Bv$. The induced instantaneous magnetization by the external field, ${\bf M}\propto \chi(0) \Bv$, has a component perpendicular to $\bJ$, which is seen rotating with respect to $\bJ$ (see Figure \ref{fig:torque-free}). The perpendicular instantaneous magnetization lags behind the grain material due to the rotation and induces internal dissipation of rotational energy, which results in the alignment of the grain angular momentum with the magnetic field.

For carbonaceous grains, paramagnetism arises from nuclear spins. Following the D-G mechanism, the characteristic time of the magnetic relaxation for nuclear magnetism is given by 
\bea
\tau_{\rm nucl-DG} &=& \frac{I_{\|}}{K_{n}(\omega)VB^{2}}=\frac{2\rho a^{2}}{5K_{n}(\omega)B^{2}},\nonumber\\
&\simeq & 1.5\times 10^6\frac{\hat{\rho}a_{-5}^{2}}{\hat{K}(\omega)\hat{B}^{2} }\,\yr,\label{eq:tau_DG}
\ena
where $\hat{K}(\omega)=K_{n}(\omega)/(10^{-13}\,{\rm s})$ with $K_{n}(\omega)$ given by Equation \eqref{eq:Kn}.

Following \cite{2016ApJ...831..159H}, to quantify the effect of nuclear magnetic relaxation on grain alignment, we calculate the ratio of the nuclear magnetic relaxation timescale to the gas damping time, 
\bea
\delta_{\rm nucl-DG}=\frac{\tau_{\rm gas}}{\tau_{\rm nucl-DG}}
\simeq 8.53\times 10^{-2} \frac{s\hat{K}(\omega)\hat{B}^{2}}{a_{-5}n_{1}T_{2}^{1/2}\Gamma_{\|}},\label{eq:delta_nucl}
\ena
which increases with the magnetic field strength and decreases with increasing gas density. One can determine the critical size for external alignment between $\bJ$ and $\Bv$ by nuclear magnetic relaxation using the condition $\delta_{\rm m,nucl}=1$, yielding

\bea
a_{\rm min,JB}^{\rm nucl-DG}=8.53\times 10^{-3}\frac{s\hat{K}(\omega)\hat{B}^{2}}{n_{1}T_{2}^{1/2}\Gamma_{\|}}\mum,\label{eq:amin_DG}
\ena
which depends on the grain angular velocity $\omega$.

\begin{figure*}
\includegraphics[width=0.5\textwidth]{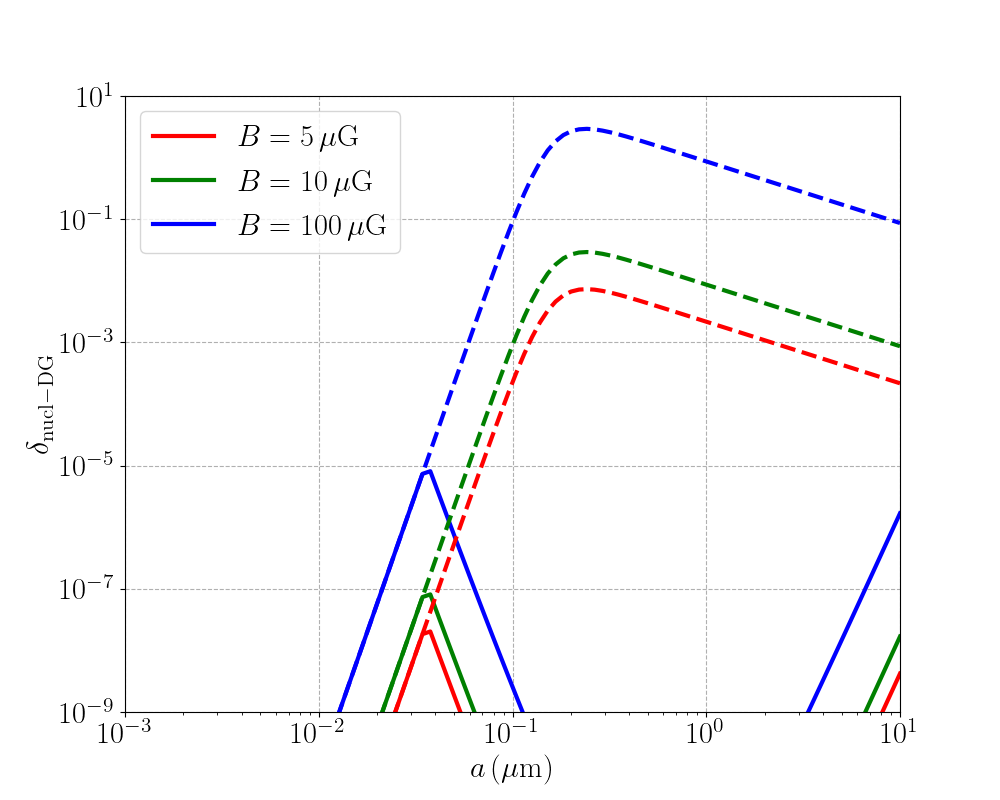}
\includegraphics[width=0.5\textwidth]{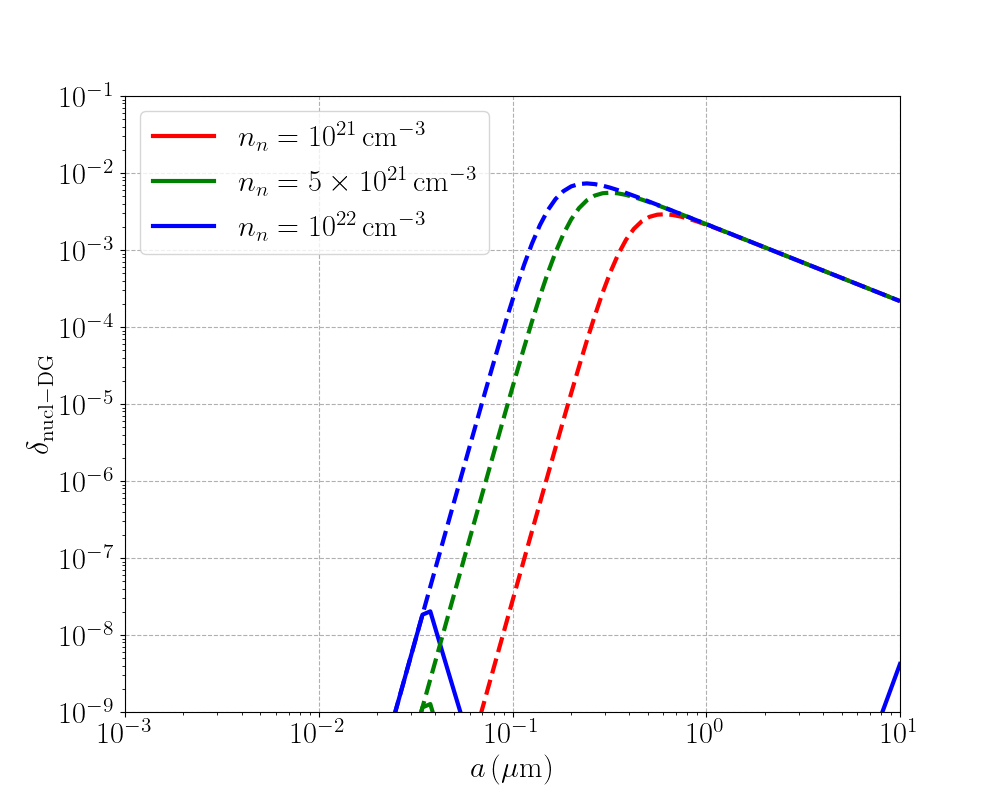}
\caption{Left panel: Magnetic parameter $\delta_{\rm nucl-DG}$ in a medium of density $n_{\rm H}=10^{1}$ cm$^{-3}$, magnetic field strength $B=5$ $\mu$G and radiation field with $U=1$, $\gamma=0.3$, and $\bar{\lambda}=1.2\,\mum$. Results are shown for grains at low-J attractors with $\Omega=\Omega_T$ (dashed lines) and high-J attractors with $\Omega={\rm max}(\Omega_{T},\Omega_{\rm RAT})$ (solid lines) assuming different values of the nucleon number density $n_n=10^{21}$ cm$^{-3}$ (red), $n_n=5\times 10^{21}$ cm$^{-3}$ (green), and $n_n=10^{22}$ cm$^{-3}$ (blue). Right panel: Same as the left panel, but for the different magnetic field strengths $B=5\,\mu$G (red), $B=10\,\mu$G (green), and $B=100\,\mu$G (blue). Results are shown for a typical value of $n_n=10^{22}$ cm$^{-3}$.}
\label{fig:delta_nn}
\end{figure*}

Figure \ref{fig:delta_nn} (left panel) shows the variation of $\delta_{\rm nucl-DG}$ as a function of the grain size for the different fraction of hydrogen protons. The value of $\delta_{\rm nucl-DG}$ increases with $n_{n}$ because of the increase of the nuclear fluctuation time $\tau_{n}$ and $K_{n}$ (see Eq.~\eqref{eq:Kn}). For the thermal rotation case, the value of $\delta_{\rm nucl-DG}$ decreases rapidly when the grain size is smaller than $a\sim 0.1\mum$ due to the strong suppression of nuclear magnetism for $\omega>2/\tau_{n}$. Beyond $a\sim 0.2\mum$, one can see $\delta_{\rm nucl-DG}$ decreases again because of the dependence on the grain size as $1/a$ (see Eq.~\eqref{eq:delta_nucl}). We present also results for the case where $\Omega={\rm max}(\Omega_T,\Omega_{\rm RAT})$ (solid lines). Note that, in this case, the suprathermal rotation by RATs is achievable only for grains of size larger than $0.05\mum$. However, the value of $\delta_{\rm nucl-DG}$ for suprathermally rotating grains is always much smaller than the one for grains with $\Omega=\Omega_T$ because of the suppression of nuclear magnetism with increasing the rotation rate. Therefore, very small carbonaceous grains (e.g., PAHs) cannot be aligned by the nuclear D-G mechanism, which is different from nanosilicates and nanoiron particles \citep{2016ApJ...821...91H}.

Figure \ref{fig:delta_nn} (right panel) shows the results for different magnetic field strengths from $B=5,10,100\mu G$, assuming the same gas density and a constant value of $n_{n}$. One can see that the strong magnetic field case can result in more efficient external alignment due to the nuclear DG mechanism if grains rotate thermally ($\Omega=\Omega_T$, dashed lined). However, as demonstrated in \cite{2016ApJ...821...91H}, grains with thermal rotation, even having superparamagnetism, cannot be aligned efficiently due to strong thermal fluctuations within the grain material. Therefore, although the nuclear DG relaxation can be effective at thermal rotation, the net alignment degree of grains is rather small less than $5\%$.


It is clear from these results that the magnetically enhanced alignment between $\bJ$ and $\Bv$ could not be achieved in the diffuse ISM given the typical values of the zero-frequency susceptibility of proton nuclei (see Eqs. \eqref{eq:chi_n} and \eqref{eq:Kn}).

\section{Application for C-rich AGB stars}\label{sec:AGB}
The envelope of AGB stars is an ideal environment to test grain alignment mechanisms due to the segregation of dust compositions and strong radiation field. Since carbon dust is thought to be formed primarily in the envelope of C-rich AGB stars, here we discuss in detail the alignment of carbon dust for the physical conditions of a typical C-rich AGB star, IRC+10216. 

\subsection{A model for the C-rich AGB envelope}
Let $\dot{M}$, $v_{\rm exp}$, and $r$ be the rate of mass loss of the AGB star, the expansion velocity of the outflow, and the distance from the center of the envelope. The hydrogen gas density is given by
\begin{eqnarray}
n_{\rm H}(r)&=&\frac{\dot{M}}{4\pi r^{2}v_{\rm exp}\mu m_{\H}}\simeq 10^{6}\left(\frac{\dot{M}}{ 10^{-5}M_{\odot}\,{\rm yr}^{-1}}\right)\nonumber\\
 &&\times\left(\frac{10\,{\rm  km~s}^{-1}}{v_{\rm exp}}\right)\left(\frac{10^{15}{\rm cm}}{r}\right)^{2}
 \,{\rm cm}^{-3},\label{eq:ngas}
 \end{eqnarray}
 where $\mu\approx 1.37$ is the mean molecular weight of the envelope.
 

The strength of the stellar radiation field is given by,
\begin{eqnarray}
U(r)&=&\frac{L_{\star}}{4\pi r^{2}c u_{\rm ISRF}}\nonumber\\
&\simeq& 1.2\times 10^{8}\left(\frac{L_{*}}{10^4 L_{\odot}}\right)\left(\frac{r}{10^{15}\cm}\right)^{-2},\label{eq:Ur}
\end{eqnarray}
where the attenuation of the stellar radiation by circumstellar dust is ignored.

In the following, we adopt the same physical parameters of IRC +10216, including the mass loss rate, the stellar temperature, and luminosity, as shown in Table 1 in \cite{2020ApJ...893..138T}. The stellar radiation field is unidirectional with $\gamma=1$, and the mean wavelength of the radiation is $\bar{\lambda}=2.5\,\mu$m (see, e.g., \citealt{2020ApJ...893..138T}). 

We adopt the toroidal model for the magnetic field strength (see e.g., \citealt{Duthu.2017,2018CoSka..48..187V})
\begin{eqnarray}
B = 1.77\times 10^{5}\left(\frac{r}{10^{14}\,{\rm cm}}\right)^{-1}\,\mu{\rm G}.
\label{eq:Bprofile}
\end{eqnarray}
Note that the magnetic field strength at the largest distance considered ($r=10^{18}$ cm) is still a few times larger than the typical ISM value of $5\,\mu$G.    


\subsection{Critical grain sizes for internal alignment}


Using the gas density and the radiation field given by Equations (\ref{eq:ngas}) and (\ref{eq:Ur}), we can estimate the critical sizes for internal alignment with both inelastic relaxation and nuclear relaxation. Note that results in this subsection are shown assuming grains are composed of HAC, and the case of $^{13}$C dominated grains will be explored also in the following subsection. 

\begin{figure*}
\includegraphics[width=0.5\textwidth]{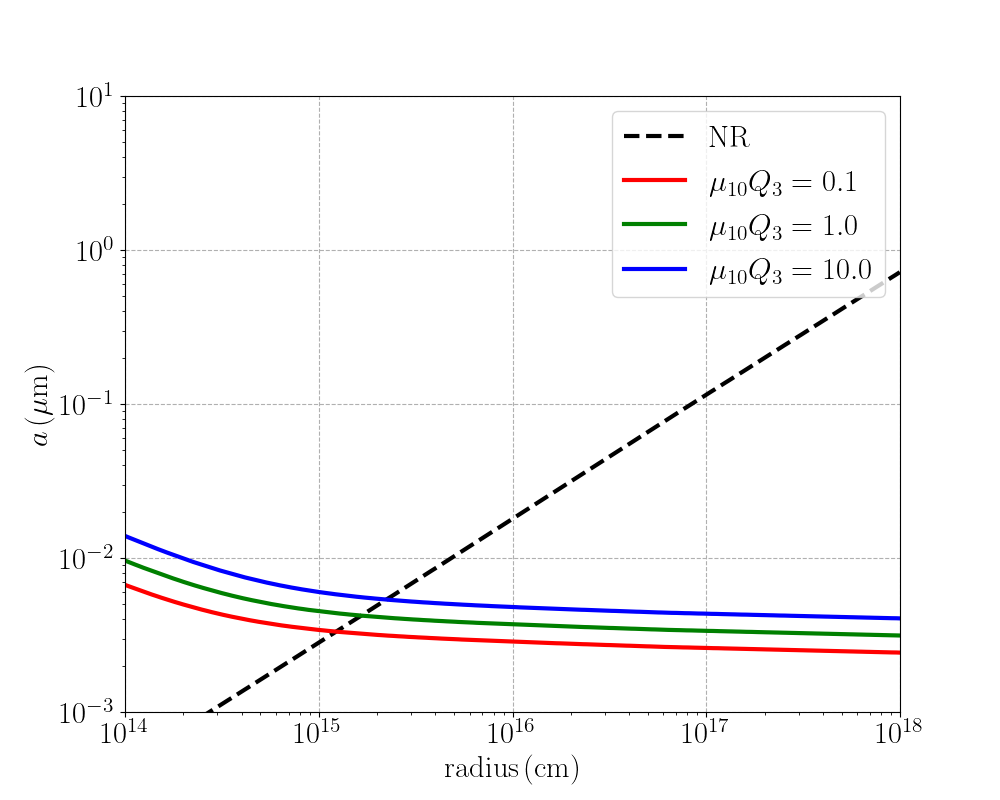}
\includegraphics[width=0.5\textwidth]{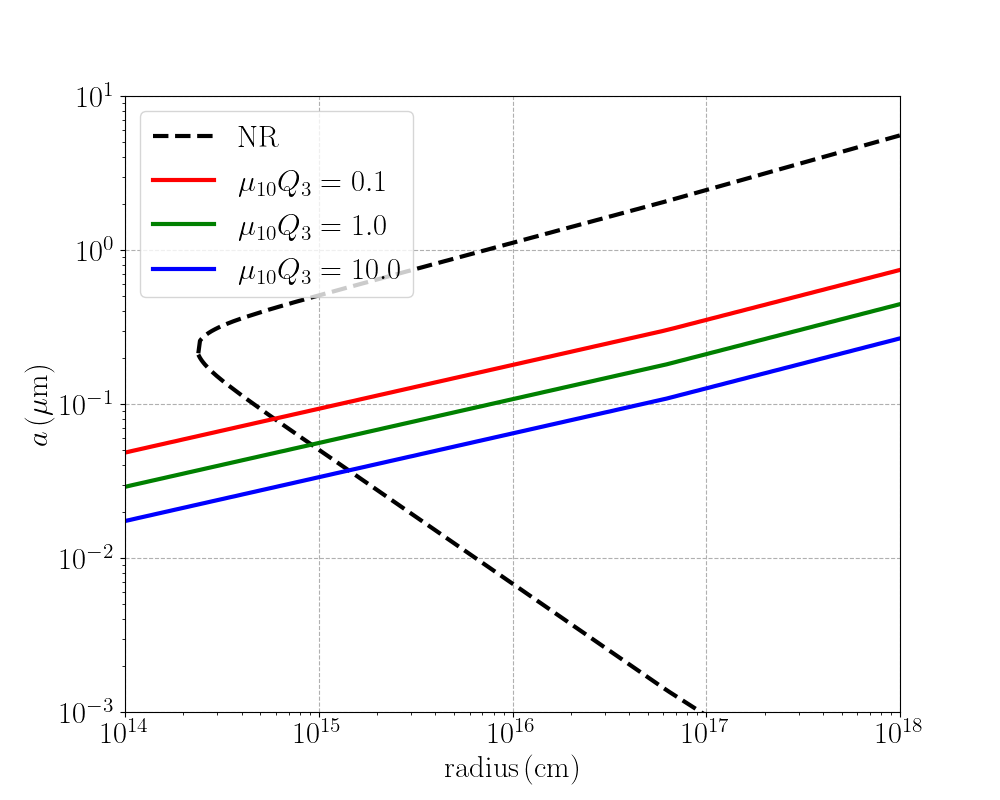}
\caption{Left panel: Critical grain sizes for efficient internal alignment for HAC grains aligned at high-J attractors via nuclear relaxation ($a^{\rm high-J}_{\rm max,aJ}(\rm NR)$, dashed line) and minimum grain size for inelastic relaxation ($a^{\rm high-J}_{\rm min,aJ}(\rm iER)$, solid lines) as a function of the radial distance from the central star for the IRC +10216. Results in the case of inelastic relaxation are shown for three different values of $\mu_{10} Q_3=0.1, 1, 10$. Right panel: Critical grain sizes for efficient internal alignment for grains aligned at low$-J$ attractor via nuclear relaxation ($a^{\rm low-J}_{\rm min,aJ}(\rm NR)$ and $a^{\rm low-J}_{\rm max,aJ}(\rm NR)$, dashed line) and inelastic relaxation ($a^{\rm low-J}_{\rm max,aJ}(\rm iER)$, solid lines).}
\label{fig:a_high-J_iER_AGB_IRC10126}
\end{figure*}

Figure \ref{fig:a_high-J_iER_AGB_IRC10126} (left panel) shows the critical sizes for internal alignment for HAC grains aligned at high-J attractors, due to nuclear relaxation ($a^{\rm high-J}_{\rm max,aJ}({\rm NR})$, dashed line) and inelastic relaxation ($a^{\rm high-J}_{\rm min,aJ}({\rm iER})$, solid line), as a function of radius from the central star. We present results in case of inelastic relaxation for $\mu_{10} Q_3=0.1$ (red), $\mu_{10} Q_3=1$ (green), and $\mu_{10} Q_3=10$ (blue). It is clear that grains at high$-J$ attractors with size greater than $\sim 10^{-2}\mum$ can have efficient internal alignment.

The right panel of Figure \ref{fig:a_high-J_iER_AGB_IRC10126} shows the similar results as the left panel but grains aligned at low-J attractors. For both nuclear relaxation and inelastic relaxation, most of the thermally rotating grains with $a\lesssim 1\,\mu$m (see $a^{\rm high-J}_{\rm min,aJ}({\rm NR}$ and $a^{\rm high-J}_{\rm max,aJ}({\rm NR}$, dashed lines) or $a\lesssim 0.03\,\mu$m in the innermost part of the envelope (see $a^{\rm low-J}_{\rm max}({\rm iER})$, solid lines) are more likely to achieve right internal alignment due to fast relaxation. Interestingly, this also means that grains with $a\gtrsim 1\,\mum$ (or $a\gtrsim 0.03\,\mu$m in the innermost part of the envelope) may have the wrong IA due to slow internal relaxation.


\subsection{Critical grain sizes for external alignment via $k$-RAT, $B$-RAT and $E$-RAT}
Here, we calculate the critical sizes for external alignment induced by RATs via $k$-RAT, $B$-RAT, or $E$-RAT. 
Note that, for calculations of $\tau_{E}$, we adopt a typical electric potential for grains induced by collisional charging of $\phi=-0.0216T_{2}\,{\rm V}$ and $v_{d}= v_{\rm exp}\simeq 15\km\s^{-1}$ (see \citealt{2020ApJ...893..138T}). 

\begin{figure*}[htpb]
\centerline{
\includegraphics[width=0.5\textwidth]{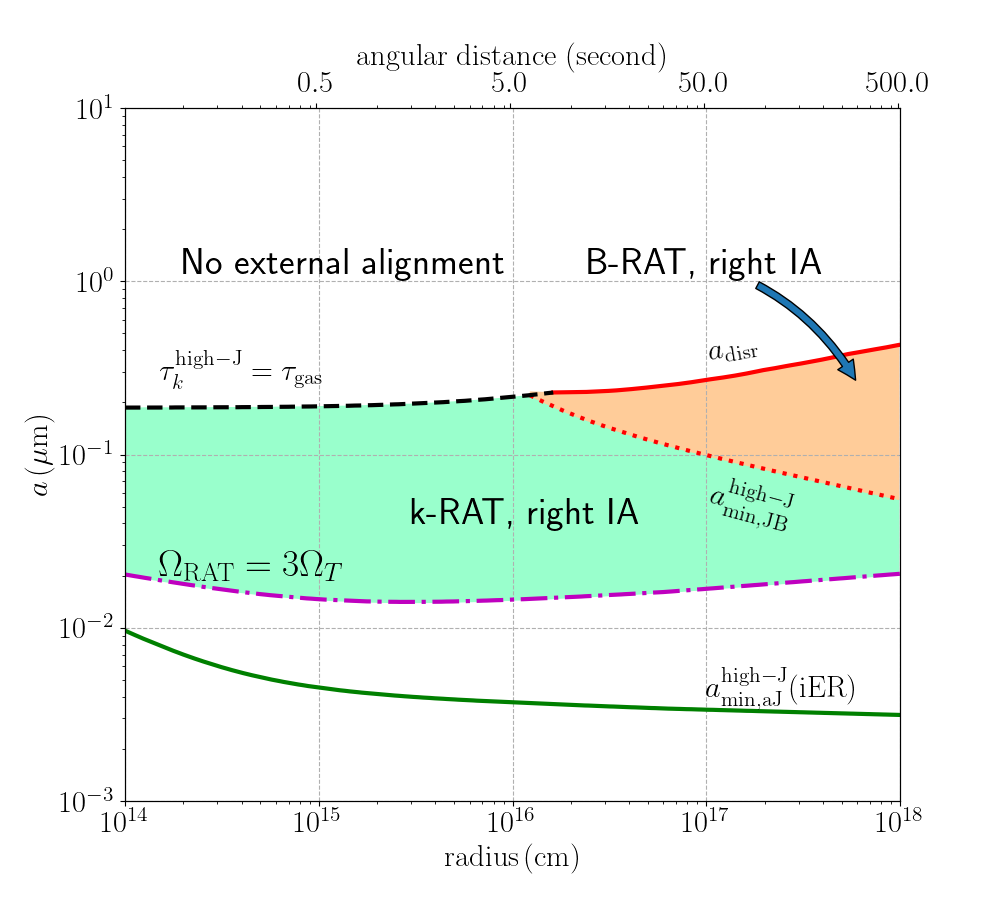}
\includegraphics[width=0.5\textwidth]{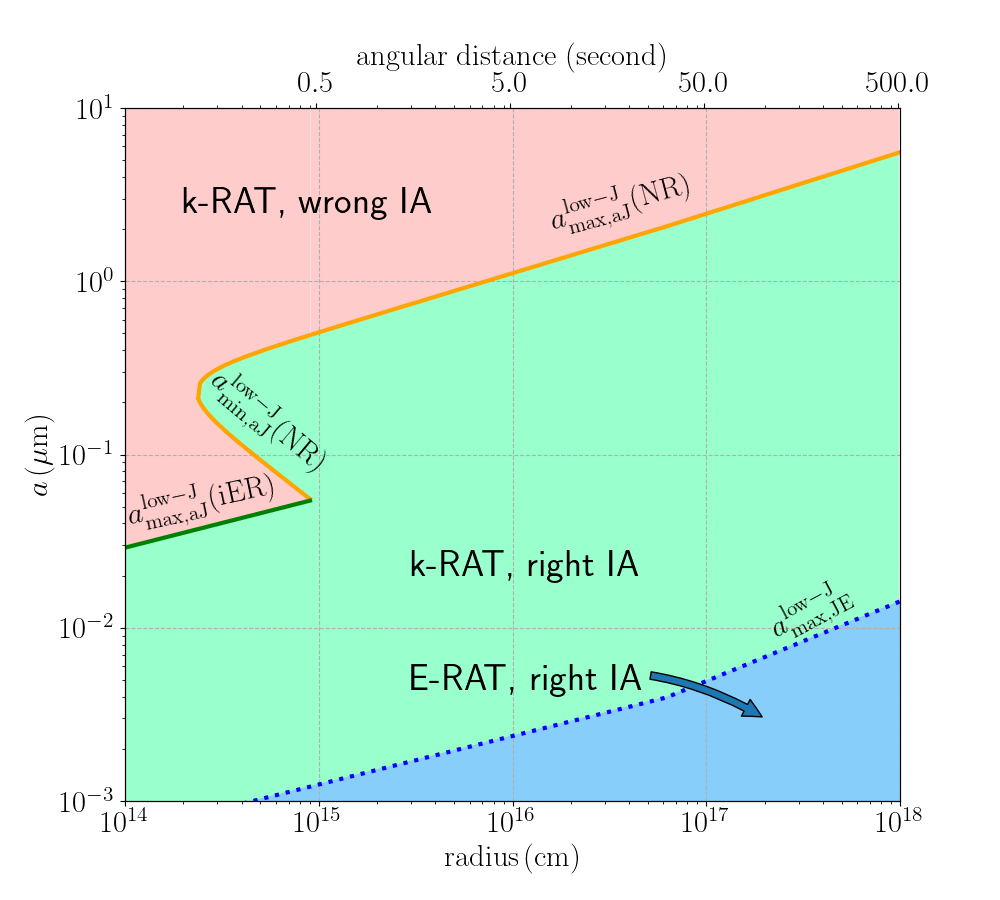}
}
\caption{Critical grain sizes for external alignment via $k$-RAT and $B$-RAT at both high$-J$ (left panel) and low$-J$ (right panel) attractors as functions of the radial distance from the center of the IRC +10216 assuming HAC grains.}
\label{fig:a_Jk_AGB_IRC10126}
\end{figure*}

\begin{figure*}[htpb]
\centerline{
\includegraphics[width=0.5\textwidth]{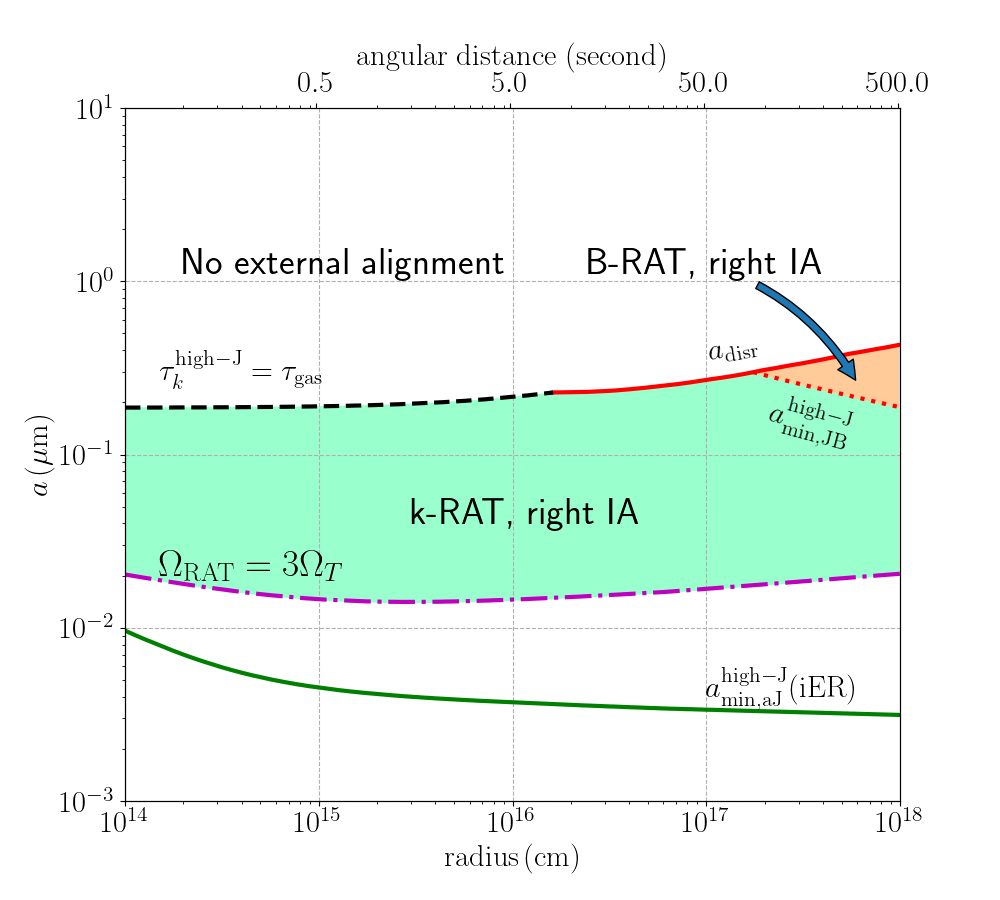}
\includegraphics[width=0.5\textwidth]{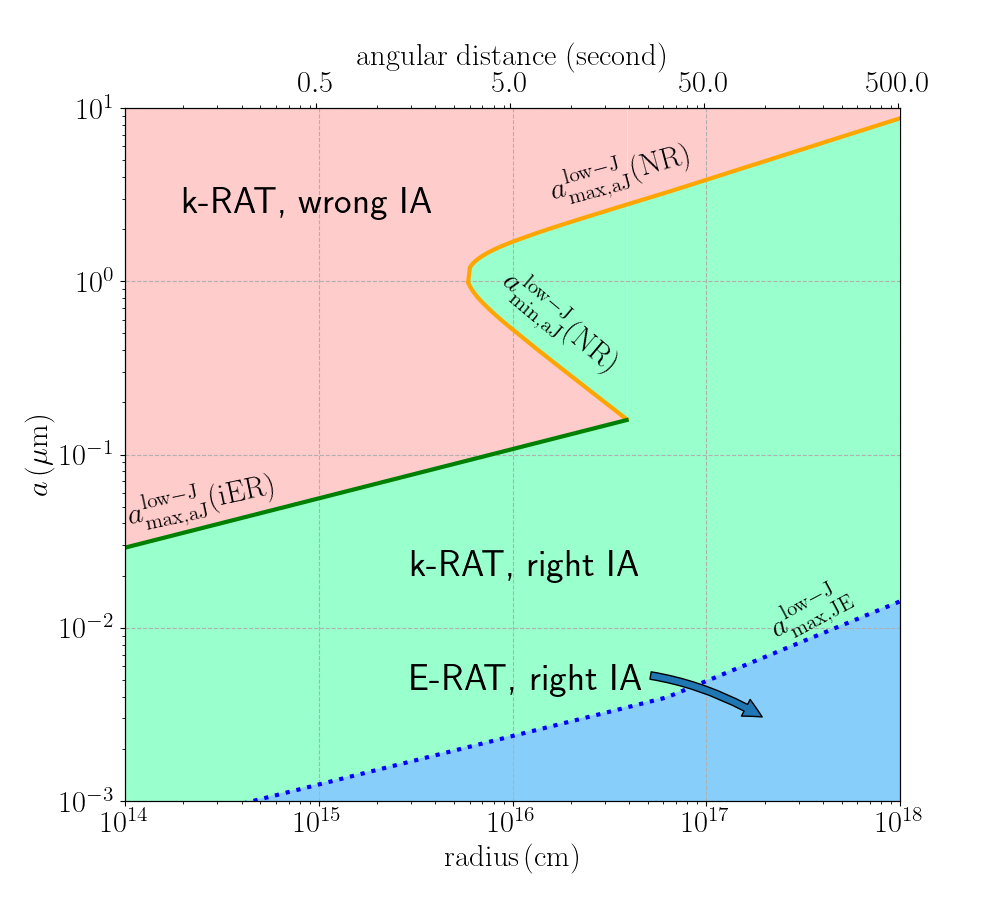}
}
\caption{Critical grain sizes for external alignment via $k$-RAT and $B$-RAT at both high$-J$ (left panel) and low$-J$ (right panel) attractors as functions of the radial distance from the center of the IRC +10216 envelope assuming $^{13}$C dominated grains.}
\label{fig:a_Jk_AGB_IRC10126_Bprofile}
\end{figure*}

We first discuss the case for HAC grains. Figure \ref{fig:a_Jk_AGB_IRC10126} shows the critical grain sizes of external alignment via $k$-RAT, $B$-RAT, and $E$-RAT for grains aligned at both high$-J$ (left panel) and  low$-J$ (right panel) attractors. The magnetic field strength is described by Eq. \ref{eq:Bprofile}). In the left panel, we also show the maximum grain size constrained by Radiative Torque Disruption (RAT-D) effect \citep{Hoang:2019bi}, denoted by $a_{\rm disr}$ (solid red line, see more details in Section \ref{sec:disrupt}) and the minimum size of RAT alignment, denoted by $a_{\rm align}$ determined by the suprathermal condition $\Omega_{\rm RAT}=3\Omega_{T}$ (dash-dotted purple line).

For the case of high$-J$ attractors (left panel), grains are mostly aligned via $k$-RAT with right IA due to efficient inelastic relaxation (see the green shaded area). Gas collisions (dashed black line) and RAT-D (solid red line), however, constrain the grain size for efficient external alignment with the radiation field to be smaller than $a\simeq 0.4\,\mum$ up to a very large distance from the central star. Interestingly, the $B$-RAT alignment with right IA can occur in the outer envelope of $r\gtrsim 10^{16}$ cm (see orange shaded area) with the grain size limited both by RAT-D and the minimum size $a^{\rm high-J}_{\rm min,JB}$ (set by requiring $\tau^{\rm high-J}_k=\tau_B$, dotted red line). 

From the right panel, one can see that in the innermost part of the envelope, grains of size $a>0.03\mum$, aligned at low-J attractors, may align via $k$-RAT with the wrong IA (see the pink-shaded area). Beyond $r\sim 5\times 10^{14}$ cm, due to the efficient nuclear relaxation, only sufficiently large grains ($a\gtrsim 1\,\mu$m) could align via $k$-RAT with the wrong IA. Smaller grains ($a\lesssim 1\,\mu$m) in this region align mostly via $k$-RAT and $E$-RAT with right IA (green- and blue-shaded areas respectively). Note also that the grain size for efficient $E$-RAT alignment with right IA, in this case, is also limited by the maximum size $a^{\rm low-J}_{\rm max,JE}$ (set by requiring $\tau^{\rm low-J}_k=\tau^{\rm low-J}_E$, dotted blue line).

We show also in Figure \ref{fig:a_Jk_AGB_IRC10126_Bprofile} the critical grain sizes of external alignment via $k$-RAT, $B$-RAT, and $E$-RAT for grains aligned at both high$-J$ (left panel) and low$-J$ (right panel) attractors for the case of $^13$C dominated grains. In this case, we set the nuclear $g$-factor $g_n=0.8$ and the density of $^{13}$C $n_n\simeq 10^{21}$ cm$^{-3}$. The latter has been chosen assuming the grain mass density of $\rho\simeq 2.2$ g/cm$^{-3}$ and the ratio $^{12}$C/$^{13}$C$\simeq 45$ as inferred from molecular line observations of IRC +10126 \citep{Pardo.2022}. 

In the figure (left panel), we show again the maximum grain size constrained by gas collisions (dashed black line) and RAT-D effect ($a_{\rm disr}$, solid red line) and the minimum size of RAT alignment ($a_{\rm align}$, dash-dotted purple line). The behavior for $^{13}$C dominated grains at high$-J$ attractors (left panel) is very similar to the case of HAC grains. Grains are mostly aligned via $k$-RAT with right IA due to efficient inelastic relaxation (green shaded area). Gas collisions (dashed black line) and RAT-D (solid blue line) limit the grain size for efficient external alignment with the radiation field to be smaller than $a\simeq 0.4\,\mum$ up to a very large distance from the central star. However, the $B$-RAT alignment with right IA can occur at a much larger distance $r\gtrsim 2\times 10^{17}$ cm (see orange shaded area) compared to the HAC case. This is because $^{13}$C dominated grains have a much smaller nuclear magnetic susceptibility and, thus, $B$-RAT is not as efficient as $k$-RAT in the inner envelope. 

It is also due to the smaller nuclear magnetic susceptibility of $^{13}$C dominated grains that nuclear relaxation for internal alignment becomes less efficient which can be seen in the right panel of Figure \ref{fig:a_Jk_AGB_IRC10126_Bprofile}. It is clear that, in this case, alignment via $k$-RAT with the wrong IA (pink-shaded area) throughout the inner envelope ($r\lesssim 10^{16}$ cm) is possible also for small grains ($a\gtrsim0.03\,\mu$m). Beyond $r\sim 10^{16}$ cm, we see gain that grains align mostly via $k$-RAT and $E$-RAT with right IA (green and blue shaded areas respectively).



\subsection{The role of the fraction of high-J attractors, $f_{\rm high-J}$}
In general, a fraction $f_{\rm high-J}$ of grain shapes can achieve suprathermal rotation so that the external alignment occurs with $\bJ$ along the radiation direction (i.e., the $k$-RAT alignment). In addition, carbon dust is expected to have right IA with the long axis perpendicular to $\kv$ (radial direction) for grains of size $a\gtrsim a^{\rm high-J}_{\rm min,aJ}({\rm iER})$. Therefore, if $f_{\rm high-J}$ is considerable, thermal emission from carbon dust is polarized with the azimuthal polarization pattern since ${\bf E}$ is oscillating along the long grain axis. On the other hand, if the majority of grains are aligned at low$-J$ attractors, internal relaxation is not efficient. In this case, internal alignment can occur with two possible states of right IA ($\ahat_{1}\| \bJ$) and wrong IA ($\ahat_{1}\perp \bJ$). Note that numerical calculations for an ensemble of grain shapes by \citep{Herranen.2021} reveal that RATs can produce $f_{\rm high-J}\sim 0.2-0.7$. Therefore, the exact value of $f_{\rm high-J}$ for carbon dust in circumstellar envelopes is uncertain.

\subsection{The effect of rotational disruption by RATs and METs}\label{sec:disrupt}
\subsubsection{Rotational disruption by RATs}
In strong radiation fields such as AGB envelopes, large grains aligned with high-J attractors can be rotationally disrupted by RATs when the centrifugal stress induced by fast rotation exceeds the grain maximum tensile strength \citep{Hoang:2019da,2019ApJ...876...13H}. This rotational mechanism is known as the RAT disruption (aka. RAT-D, \citealt{2020Galax...8...52H}). 
 The critical angular velocity for the RAT-D is given by \cite{Hoang:2019da}
\begin{eqnarray}
\Omega_{\rm disr}\simeq 1.1 \times 10^{10}a_{-5}^{-1}\hat{\rho}^{-1/2}\left(\frac{S_{\rm max}}{10^{10}\, {\rm erg cm^{-3}}}\right)^{1/2}\,{\rm rad\, s}^{-1},
\end{eqnarray}
where $S_{\rm max}$ is the maximum tensile strength of dust material. 
The critical size for the RAT-D is determined by setting the maximum angular velocity spun-up by RATs, $\Omega_{\rm RAT}$ (Eq. \ref{eq:omega_RAT1}), equal to $\Omega_{\rm disr}$.

\begin{figure}
\includegraphics[width=0.5\textwidth]{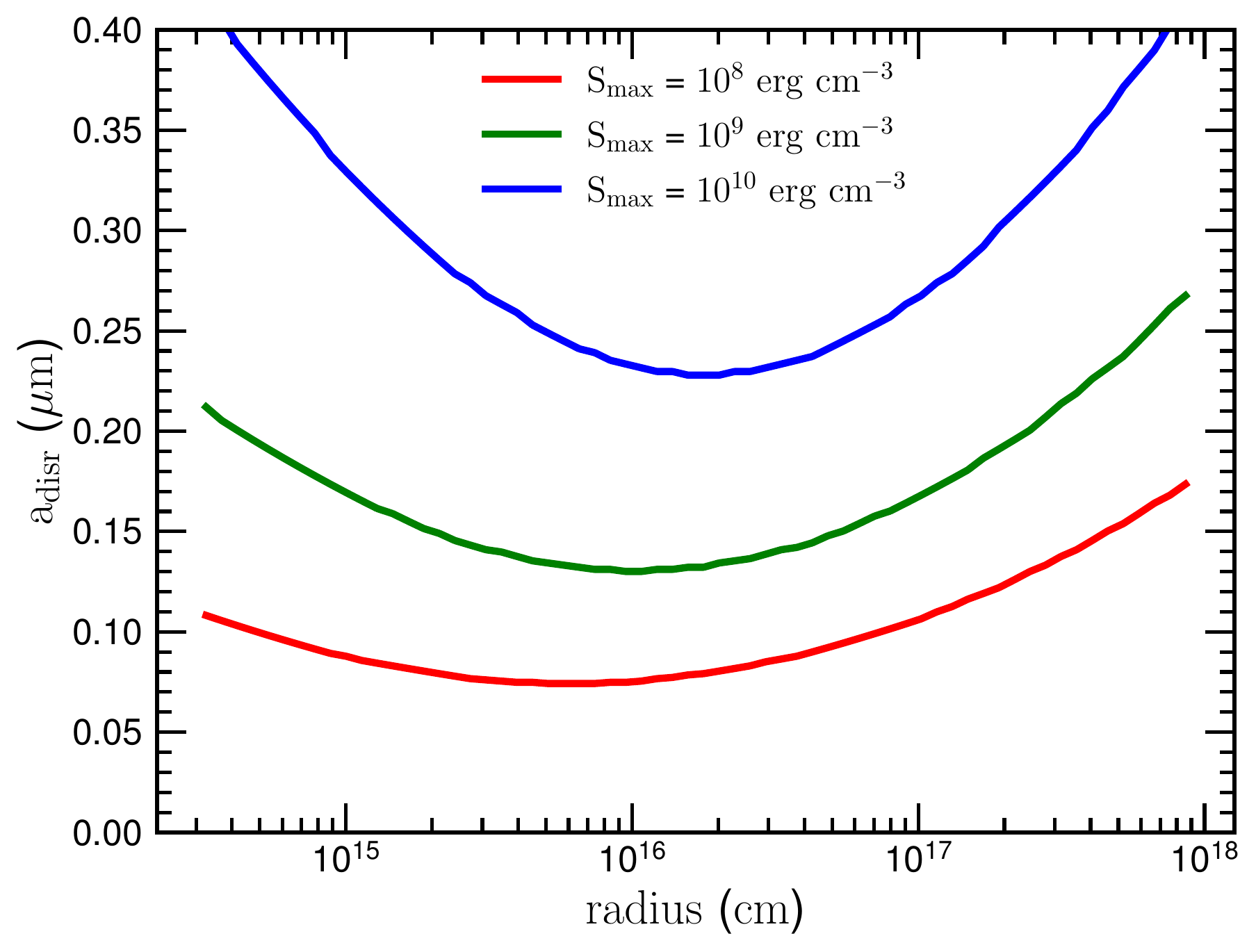}
\caption{Disruption size ($a_{\rm disr}$) as a function of radius from the IRC +10216 for different assumptions of the maximum tensile strength ($S_{\rm max}$). The initial maximum grain size is adopted as 10$\,\mu$m. Close to the central IRC +10216, the rotation damping is very efficient owing to an extremely dense and hot gas, making a low RAT-D efficiency. Farther in the envelope, in which $n_{\rm H}$ and $T_{\rm gas}$ substantially drop, the rotation damping becomes less efficient and hence the disruption is more efficient. One can see that smaller grains are disrupted for lower $S_{\rm max}$. Further decreasing of $n_{\rm H}$ and $T_{\rm gas}$, a specific conditions with $F_{\rm IR}\gg 1$ and $U\gg 1$ can be satisfied, $a_{\rm disr}$ could be computed analytically. In this case, $a_{\rm disr}$ tends to increase due to lower radiation intensity coming from the central source. See \cite{2020ApJ...893..138T} for more details.}
\label{fig:adisr_IRC10216}
\end{figure}

A detailed modeling of the RAT-D effect for the envelope of AGB stars in \cite{2020ApJ...893..138T} shows that large carbonaceous grains can be disrupted efficiently. The disruption size, $a_{\rm disr}$, changes across the envelope, as shown in Figure \ref{fig:adisr_IRC10216}. Thus, a fraction $f_{\rm high-J}$ of large grains with size $a>a_{\rm disr}$ that are aligned at high$-J$ attractors will be disrupted into smaller ones. On the other hand, the fraction $1-f_{\rm high-J}$ of large grains of $a>a_{\rm disr}$ that are aligned at low$-J$ attractors cannot be disrupted due to slow rotation. This has an important implication for dust polarization.

For the grains at low$-J$ attractors, slow internal relaxation occurs for large grains of $a>a_{\min,aJ}^{\rm low-J}$. Such large grains with slow internal relaxation can have the right IA with a fraction ($f_{\rm right-IA}$) and the fraction $(1-f_{\rm right-IA})$ grains have the wrong IA. Moreover, grains at low$-J$ attractors are predicted to have the $k$-RAT alignment instead of $B$-RAT due to the increase of the radiative precession rate with the radiation strength. 

\begin{figure}
\includegraphics[width=0.5\textwidth]{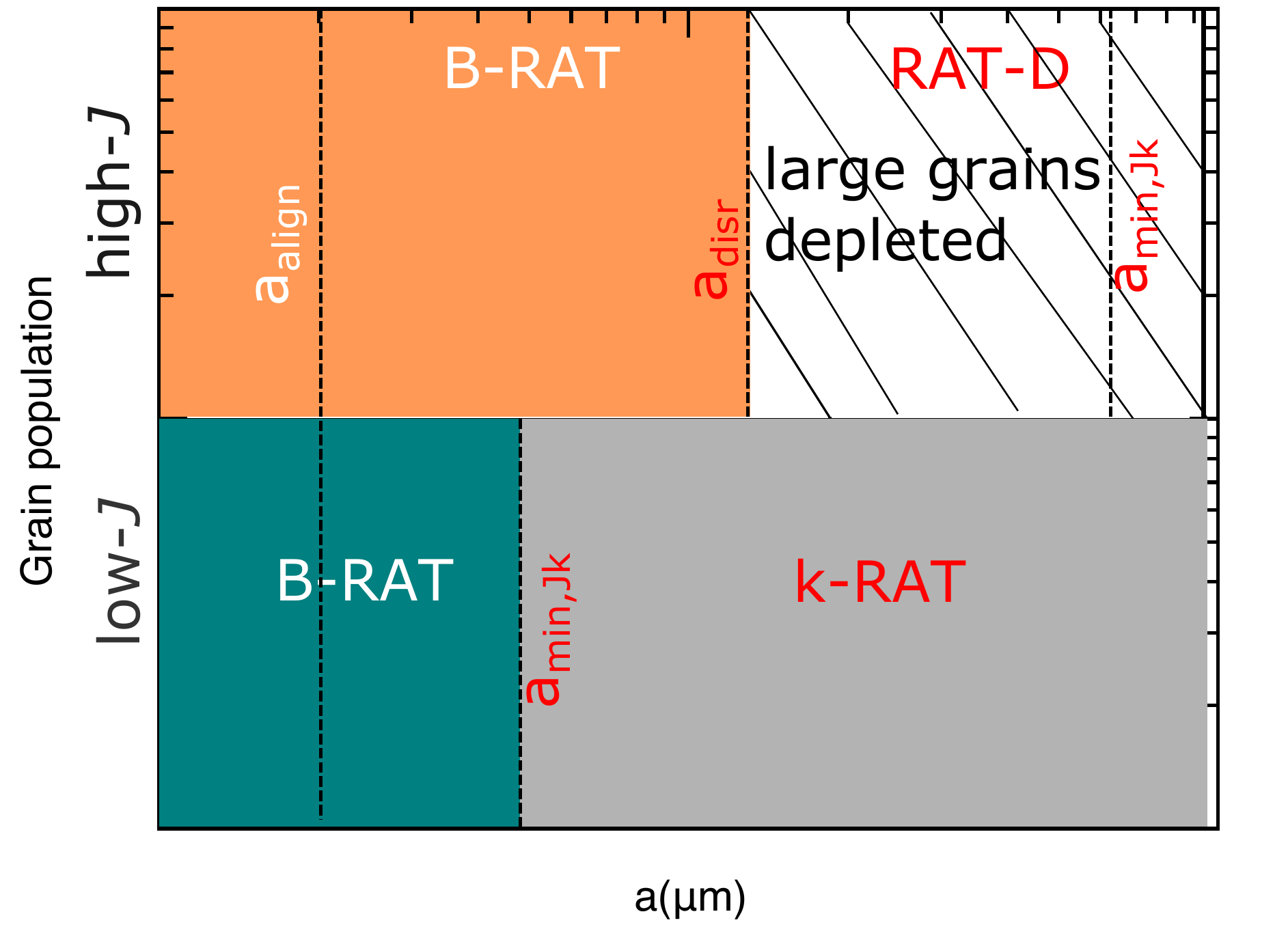}
\caption{The effect of RAT-D on the population of grains at low$-J$ and high$-J$ attractors for carbonaceous grains subject to a strong radiation field. The RAT-D removes large grains at high$-J$ attractor of $a>a_{\rm disr}$, so that larger grains are aligned at low$-J$ attractors. Large grains at low$-J$ attractors tend to experience the $k$-RAT alignment (gray area) due to fast radiative precession. Grains experience the $B$-RAT alignment for low$-J$ (green area) and high$-J$ attractors (orange area).}
\label{fig:RATD_lowJ_highJ}
\end{figure}

Figure \ref{fig:RATD_lowJ_highJ} illustrates the redistribution of grains at low$-J$ and high$-J$ attractors due to the RAT-D effect. Due to the RAT-D, grains larger than $a_{\rm disr}$ rotate thermally at low$-J$ attractors, i.e., $f_{\rm high-J}(a>a_{\rm disr})=0$ and $f_{\rm low-J}(a>a_{\rm disr})=1$. In a strong radiation field, the RAT-D can be efficient in reducing large grains at high$-J$ attractors. Therefore, the dust polarization at long wavelengths will be dominated by grains at low$-J$ attractors. As a result, the dust polarization is reduced significantly by RAT-D. Moreover, because grains at low$-J$ attractors are predicted to have the $k$-RAT alignment instead of $B$-RAT, the polarization pattern depends on the radiation direction and the fraction of grains with right IA ($f_{\rm right-IA}$). For the circumstellar envelope, the radiation direction is radial, one then has the radial polarization pattern if the wrong IA dominates, and the azimuthal pattern if the right IA dominates.

\subsubsection{Rotational disruption by METs}
 One possibility is that mechanical torques (METs) induced by the gas flow \citep{2018ApJ...852..129H} could spin up all small grains at high-J attractors to the critical angular velocity required for disruption \citep{Hoang:2018es}. The maximum angular velocity of grains spun-up by METs can be written as \citep{Hoang.20220o}
\begin{eqnarray}
\Omega_{\rm MET}=4.5\times 10^{6}\left(\frac{v_d}{v_T}\right)^2\frac{sT_{\gas,1}^{1/2}}{a_{-5}}\frac{Q_{\rm spinup,-3}}{\Gamma_{\|}}~\rad\s^{-1},
\label{eq:omega_MET}
\end{eqnarray}
where $v_d$ is the drift speed of grains, $v_T=\sqrt{2k_BT_{\rm gas}/m_{\rm H}}$ is the thermal speed of the gas, $Q_{\rm spinup,-3}=Q_{\rm spinup}/10^{-3}$ with $Q_{\rm spinup}$ is the spin-up efficiency for METs, and $\Gamma_{\|}$ is the geometrical factor as introduced in Eq. \eqref{eq:tgas}. In the following, we shall choose $Q_{\rm spinup,-3}=1$ and $\Gamma_{\|}=1$.

\begin{figure}[htpb]
\includegraphics[width=0.5\textwidth]{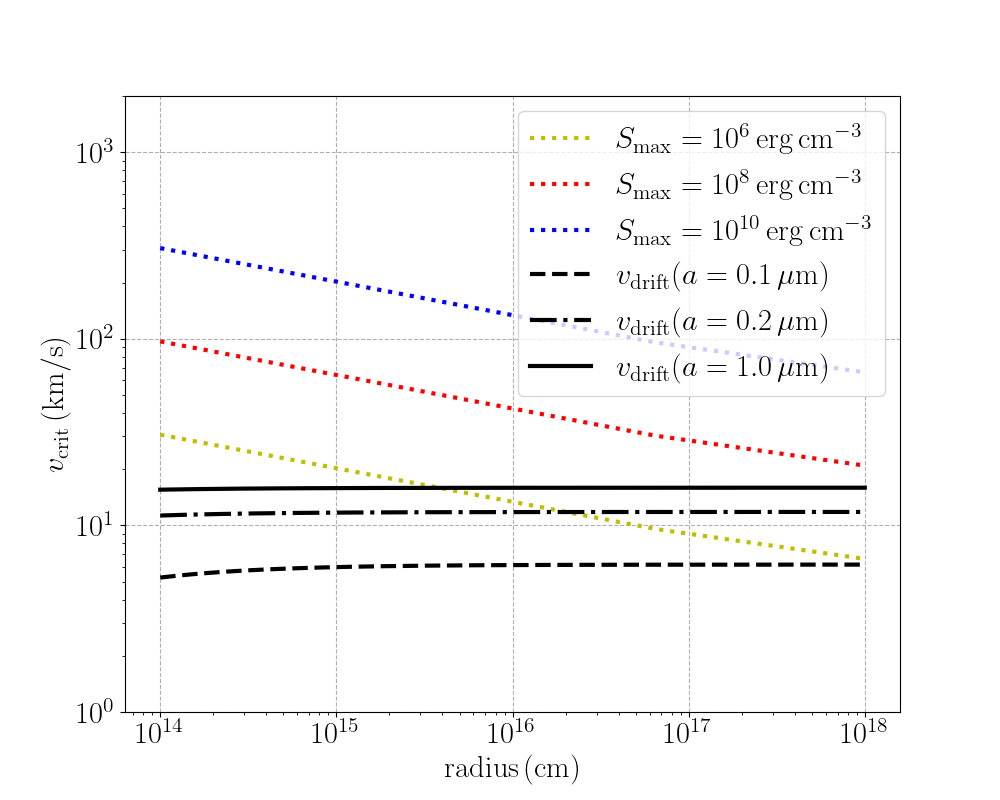}
\caption{Critical speed for MET-D with $S_{\rm max}=10^{6}$ erg cm$^{-3}$ (yellow), $S_{\rm max}=10^{8}$ erg cm$^{-3}$ (red), and $S_{\rm max}=10^{10}$ erg cm$^{-3}$ (blue). The black line represents the expansion speed of the IRC +10216 envelope.}
\label{fg:vcrit_METD}
\end{figure}

The disruption by MET-D occurs when $\Omega_{\rm MET}=\Omega_{\rm disr}$. In Figure \ref{fg:vcrit_METD} we present the critical speed for MET-D at different values of $S_{\rm max}$ denoted as $v_{\rm crit}$ which is obtained by requiring $\Omega_{\rm disr}=\Omega_{\rm MET}$. The expected drift speed of grains with size $a=0.1\,\mum$ (dashed), $a=0.2\,\mum$ (dash-dotted), and $a=1\,\mum$ (solid) is also overlaid for comparison (note that the drift speed has been obtained as in \citealt{2020ApJ...893..138T}). 

It is clear that varying $S_{\rm max}$ by 4 orders of magnitude only result in a change by roughly 1 order of magnitude in $v_{\rm crit}$. More importantly, MET-D is expected to be efficient for $v_{\rm crit}<v_{d}$ which should be more easily achievable in the outer envelope ($r>10^{16}$ cm) and also for grains with low $S_{\rm max}$. This means that, for both RAT-D and MET-D, small values of $S_{\rm max}$ in the outer envelope are needed. More quantitative investigations are required for better understanding of the polarization pattern of IRC +10216. It is interesting, however, to stress that if qualitative explanations discussed above could be assumed to be correct, one might also turn the argument around and employ the polarization data to provide constraints for the tensile strength or more generally mechanical properties of grains in the outer envelope.

\subsection{Implications for polarization observations toward IRC +10216 by SOFIA/HAWC+}
Using SOFIA/HAWC+, \cite{Andersson.2022} first reported the radial polarization pattern at $254 \mum$ from IRC +10216 and the polarization degree is rather low of $p<1\%$. As discussed in \cite{Andersson.2022}, such a polarization pattern is inconsistent with the $B$-RAT alignment. The authors suggested that carbonaceous grains are aligned by $k$-RAT with the wrong IA due to inefficient internal relaxation of carbonaceous grains. However, our detailed studies here reveal that carbonaceous grains can have the wrong IA only if they are aligned at low$-J$ attractors. Carbonaceous can still have the right IA due to fast inelastic relaxation if they rotate suprathermally at high-J attractors. In particular, we show that the $k$-RAT alignment with high$-J$ attractor is unstable because the radiative precession is much slower than gas collisions. 

Moreover, $^{13}$C dominated grains that can be rapidly aligned at high$-J$ attractors on a timescale shorter than $\tau_{\rm gas}$, which are expected to produce high polarization with azimuthal polarization, RAT-D can disrupt grains larger than $a_{\rm disr}$ into smaller fragments (see Figure \ref{fig:adisr_IRC10216}), leaving all larger grains on low-J attractors. We notice however, that grains with $a\gtrsim 0.08-0.4\,\mum$ are more likely to have $k$-RAT with wrong IA in the inner envelope which might provide explanations for the observations of polarization pattern from \cite{Andersson.2022}. Beyond ($r\gtrsim2\times 10^{17}$ cm), grains of size $a\sim 0.1-0.5\mum$ can be efficiently aligned via $B$-RAT at high-J attractors due to the reduction of RAT-D and gas density. This transition from the $k$-RAT to $B$-RAT alignment can change the dust polarization from the radial pattern to dipole maps. Interestingly, JCMT observations at 850$\mum$ seem to reveal such a feature ($B$-G Andersson, in private communication). 

\section{Discussion}\label{sec:discuss}
We study the alignment of carbonaceous grains using the RAT alignment paradigm by considering various effects, including nuclear relaxation, inelastic relaxation, nuclear Davis-Greenstein relaxation. Below, we discuss the implications for alignment of carbon dust in the ISM and circumstellar envelope of carbon-rich evolved stars.

\subsection{Alignment of interstellar carbonaceous grains within the RAT paradigm}
\subsubsection{Efficient internal alignment of carbon dust}
Using the RAT paradigm, we first quantified the critical sizes for internal alignment induced by nuclear and inelastic relaxations. For grains at low$-J$ attractors with thermal rotation, both nuclear relaxation and inelastic relaxation are efficient, leading to the internal alignment of carbonaceous grains. For grains aligned at high$-J$ attractors, we showed that both nuclear relaxation and inelastic relaxation are significantly enhanced, which induces the internal alignment of very large grains (see Eq.~\eqref{eq:amax_aJ_RAT}). The range of sizes for carbonaceous grains that have efficient internal relaxation depends on the local gas conditions and radiation field. For the diffuse ISM and translucent clouds ($n_{\rm H}<10^{3}\cm^{-3}$), carbonaceous grains can have efficient internal alignment due to inelastic relaxation.

\subsubsection{External alignment: competition of $k$-RAT, $B$-RAT vs. $E$-RAT}
We studied the external alignment within the RAT paradigm, assuming grains aligned at low$-J$ and high$-J$ attractors, and quantified the conditions for grain alignment along the radiation direction ($k$-RAT) and along the magnetic field ($B$-RAT). We find that both HAC and graphite grains have $k$-RAT alignment when they are aligned at low$-J$ attractors. However, graphite grains cannot stably align at high$-J$ attractor because radiative precession is much slower than gas collisions. Moreover, HAC grains can still be aligned via $B$-RAT when aligned at high$-J$ attractors due to the suppression of radiative precession. However, the $B$-RAT alignment of HAC grains is unstable because gas collisions induce faster randomization than the Larmor precession due to low nuclear magnetic susceptibility. Therefore, neither the radiation nor magnetic field can be the stable axis of grain alignment of carbon dust in the space if grains are assumed to be aligned at high$-J$ attractors by RATs. 

In particular, we find the in the effect of $E$-RAT is dominant over $k$-RAT and $B$-RAT for grains aligned at low-J. For grains aligned at high-J, $E$-RAT is dominant over $B$-RAT for graphite or HAC grains of small sizes (see Figure \ref{fig:tau_kBE}). 

We also quantified the effect of nuclear magnetism on the external alignment via Davis-Greenstein paramagnetic relaxation and found that the nuclear D-G effect is not efficient due to the suppression of nuclear magnetic susceptibility for fast rotation. Thus, the external alignment of carbon dust and the resulting dust polarization is only determined by RATs. Thus, the degree of dust polarization is then determined by the fraction of grains that can align with high$-J$ attractors by RATs.

\subsubsection{Grain charging and E-RAT alignment}
Grain alignment with the electric field relies on the interaction of the electric dipole moment, which depends on the grain charge. The grain charge depends on dust and gas properties, as well as the ambient radiation field. As a result, the efficiency of E-RAT also varies with the local environments and grain sizes.

In the envelope of cold AGB stars, stellar UV radiation is lacking, and collisional charging is dominant over photoelectric effect for grain charge and electric dipole moment. Our results in Figure \ref{fig:a_Jk_AGB_IRC10126_Bprofile} show that the E-RAT alignment is inefficient for large grains. However, interstellar UV radiation can penetrate into the inner envelope through a clumpy structure (e.g., \citealt{Ortiz.2016,Decin.2010} or UV radiation can be produced by a hot companion star (e.g., a white dwarf), which induces photoelectric charging. As a result, the efficiency of E-RAT alignment increases with the UV radiation strength but decreases with the electron density due to the dependence of the grain charge on $G$ and $n_{e}$ (see Section \ref{sec:charge}). Detailed modeling of grain charge and grain alignment for carbon-rich AGB stars is beyond the scope of this paper.



\subsection{The oscillation between $k$-RAT to $B$-RAT alignment}

We identified an interesting effect related to alignment of HAC, namely the oscillation between $k$-RAT to $B$-RAT. Suppose HAC grains suddenly are beamed by a radiation flash (e.g., from a SN). RATs initially are rotating slowly and have the $k$-RAT alignment. As soon as radiation flash rapidly drives grains to high$-J$ attractors (i.e., fast alignment), the grain angular momentum increases, resulting in the decrease of radiative precession rate (see Eq.~\eqref{eq:tauk}). When the radiative precession becomes slower than the Larmor precession, the axis of grain alignment changes from $\kv$ to $\Bv$. If the magnetic field makes a large angle ($\psi$) with $\kv$, the maximum angular momentum of grains aligned with $\Bv$ is $\Omega_{\max,JB}=\Omega_{\rm RAT}\cos\psi$. As a result, the radiative precession rate can be increased again and becomes larger than the Larmor precession. As a result, the axis of grain alignment goes back to $\kv$. This process occurs again and induces the oscillation between $k$-RAT and $B$-RAT, leading to the oscillation of dust polarization vectors from being perpendicular to $\kv$ to $\Bv$. The change in the alignment axis can change the observed dust polarization pattern, which can be tested with time-domain astrophysics of supernovae and gamma-ray bursts. 

\subsection{Origin of non-detection of polarized 3.4 $\mum$ C-H features}
HAC is usually suggested to be an important component of solid carbon in the ISM. However, the non-detection of polarized 3.4 $\mum$ C-H features by \cite{2006ApJ...651..268C} suggests that carbon grains are weakly aligned or have spherical shapes. Based on the RAT paradigm, our study here based on the characteristic timescales showed that HAC grains can experience efficient internal alignment due to nuclear relaxation and inelastic relaxation, but the external alignment of large HAC grains is complex, including $E$-RAT for grains aligned at low$-J$ attractors and $B$-RAT for grains of size $a>0.05\mum$ at high-J attractors (see Figure \ref{fig:tau_kBE}). On the other hand, very large grains of size $a>10\mum$ can align via $k$-RAT at low$-J$ attractors because radiative precession is much slower than gas collisions for grains at high-J attractors that destabilizes the $k$-RAT alignment. Because grains rotating thermally at low$-J$ attractors have a rather low degree of alignment due to internal thermal fluctuations and gas collisions \citep{2016ApJ...831..159H,2016ApJ...821...91H}), the polarization induced by aligned carbonaceous grains of size $a<0.05\mum$ is expected to be small. Note that small HAC grains of $a<a_{\rm align})$ (see Eq. \ref{eq:aali_RAT}) are inefficiently aligned with a degree of less than $5\%$ (\citealt{2016ApJ...831..159H,2016ApJ...821...91H}) and thus produce negligible polarization. This can explain the non-detection of polarized 3.4 $\mum$ C-H features by \cite{2006ApJ...651..268C}.

Moreover, if large HAC grains of $a>0.05\mum$ are present in the ISM, the efficient internal alignment and $B$-RAT alignment can produce considerable polarization, which is inconsistent with the non-detection of polarized 3.4 C-H feature. Therefore, our results suggest that interstellar HAC are likely to present in small grains or in the thin mantle covering silicate cores.

\subsection{Constraining grain physics and composition in C-rich AGB stars with dust polarization}
Observations of dust polarization toward C-rich AGB stars provide a direct constraint on the alignment mechanisms of carbon dust. Our theoretical calculations show that carbon grains in the envelope of IRC+10216 aligned at high-J attractors can have efficient internal relaxation by inelastic relaxation, and the external alignment occurs via $k$-RAT. Therefore, their thermal emission is polarized with the azimuthal polarization. Nevertheless, grains aligned at high-J attractors can be reduced by the RAT-D mechanism, leaving only large grains aligned at low-J attractors. As a result, the thermal polarization may have a radial pattern due to the wrong IA by slow inelastic relaxation. This may explain the radial polarization pattern observed by \cite{Andersson.2022}. A detailed modeling is beyond the scope of this paper and will be presented in our follow-up study.

Our numerical studies based on grain alignment physics in Section \ref{sec:AGB} shows that the dependence of external alignment on grain composition due to magnetic susceptibility. For HAC with nuclear magnetism caused by hydrogen, the transition from $k$-RAT to $B$-RAT for grains at high-$J$ attractors occurs at a much smaller radius $r\gtrsim 10^{16}$ cm compared to the case of $C^{13}$ dominated grains. Also, in this case, $k$-RAT with wrong IA for grains at low-$J$ attractors in the inner envelope ($r\lesssim 10^{16}$ cm) is only efficient for sufficiently large grains ($a\gtrsim 1\,\mu$m). In reality, grains in envelopes of C-rich AGB stars might have a mixed composition, e.g. grains could comprise primarily $^{13}C$ in the inner envelope but become more protonated and become HAC grains at a sufficiently large distance from the central stars. Gradual enrichment of carbonaceous grains by iron when expanding would also increase the grain magnetic susceptibility and thus induces B-RAT at the outer envelope. Therefore, observations of transitions in polarization patterns from the inner to outer envelope of evolved stars could open up a possibility to constrain the composition of carbonaceous grains.    

\section{Summary}\label{sec:concl}
We study in detail internal alignment (IA) and external alignment of carbonaceous grains, including hydrogenated amorphous carbons (HAC) and graphite, using the RAdiative Torque (RAT) paradigm with grains aligned at low-J and high-J attractors. We first study the typical ISM conditions and then apply our theory for the circumstellar envelope of C-rich AGB stars. Our main results are summarized as follows:

\begin{enumerate}
\item The nuclear magnetic susceptibility of HAC due to hydrogen protons is reduced significantly for grains rotating suprathermally at high-J attractors. This results in slow nuclear relaxation at high-J attractors and inefficient alignment by paramagnetic relaxation. However, nuclear relaxation is efficient for HAC grains at low-J attractors, leading to the right IA with the long axis perpendicular to the grain angular momentum.

\item Internal relaxation by inelastic effect is faster than the gas damping and becomes stronger for grains aligned at high$-J$ attractors with suprathermal rotation. Thus, suprathermally rotating carbonaceous grains can have right IA by inelastic relaxation.

\item For external alignment, we find that graphite grains can only be aligned along the radiation direction at low$-J$ attractors due to their lack of nuclear paramagnetism, but cannot align at high$-J$ attractors due to the slow radiative precession compared to gas collisions at suprathermal rotation.

\item HAC grains with nuclear magnetic moment have the Larmor precession faster than the radiative precession at high$-J$ attractors, resulting in $B$-RAT alignment. However, the $B$-RAT alignment is unstable because the Larmor precession is still slower than gas collisions due to the low susceptibility.

\item We suggest that interstellar HAC grains are likely to present in small sizes which are inefficiently aligned. This can explain the non-detection of polarization of $3.4\mum$ C-H features.

\item We applied our analysis for grain alignment in the circumstellar envelope of C-rich star, IRC+10216. We find that grains can be aligned via k-RAT in most of the envelope, but B-RAT alignment can occur in the outer region, which causes the change in the polarization pattern in the outer envelope. This prediction would be tested with observations.

\item In circumstellar envelopes, the axis of grain alignment can change between $\kv$ and $\Bv$ due to the variation of local radiation field, gas density, and magnetic fields, which results in the oscillation between $k$-RAT and $B$-RAT alignment. The polarization vectors change from being perpendicular to $\kv$ to $\Bv$. 

\item Moreover, the radiative torque disruption effect can remove large grains aligned at high$-J$ attractors, producing a population of large grains aligned at low$-J$ attractors only. Such slowly rotating grains tend to have $k$-RAT alignment instead of $B$-RAT due to fast radiative precession, which induces a radial polarization pattern for grains with wrong IA. The latter could explain the observations by SOFIA/HAWC+ toward IRC+10216.

\end{enumerate}

\begin{acknowledgments}
T.H. acknowledges the support from the National Research Foundation of Korea (NRF) grant funded by the Korea government (MSIT) (2019R1A2C1087045). This work was partly supported by a grant from the Simons Foundation to IFIRSE, ICISE (916424, N.H.). We would like to thank the ICISE staff for their enthusiastic support.
\end{acknowledgments}

\bibliography{ms.bbl}

\appendix

\begin{figure*}[htpb]
\includegraphics[width=0.33\textwidth]{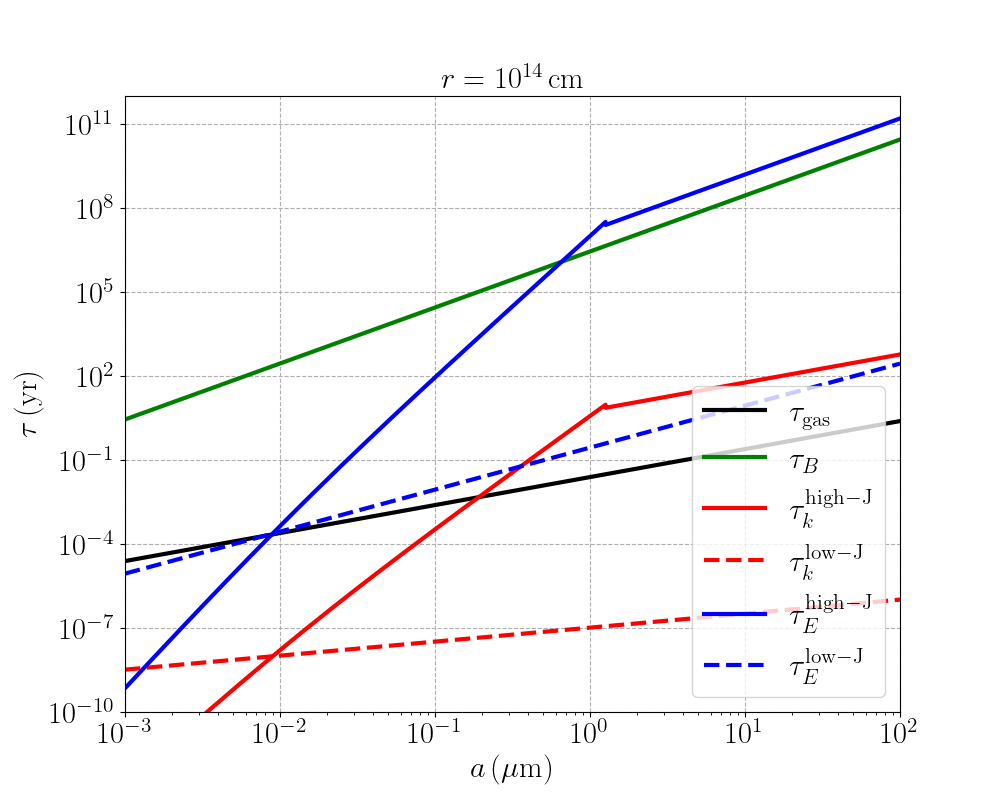}
\includegraphics[width=0.33\textwidth]{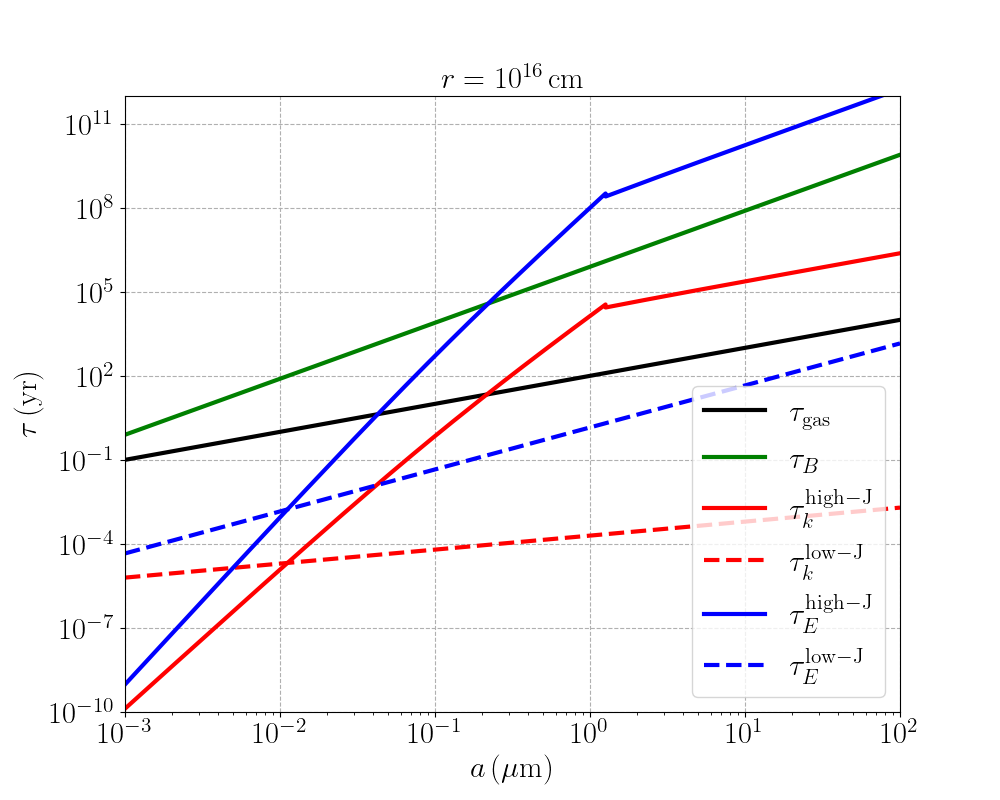}
\includegraphics[width=0.33\textwidth]{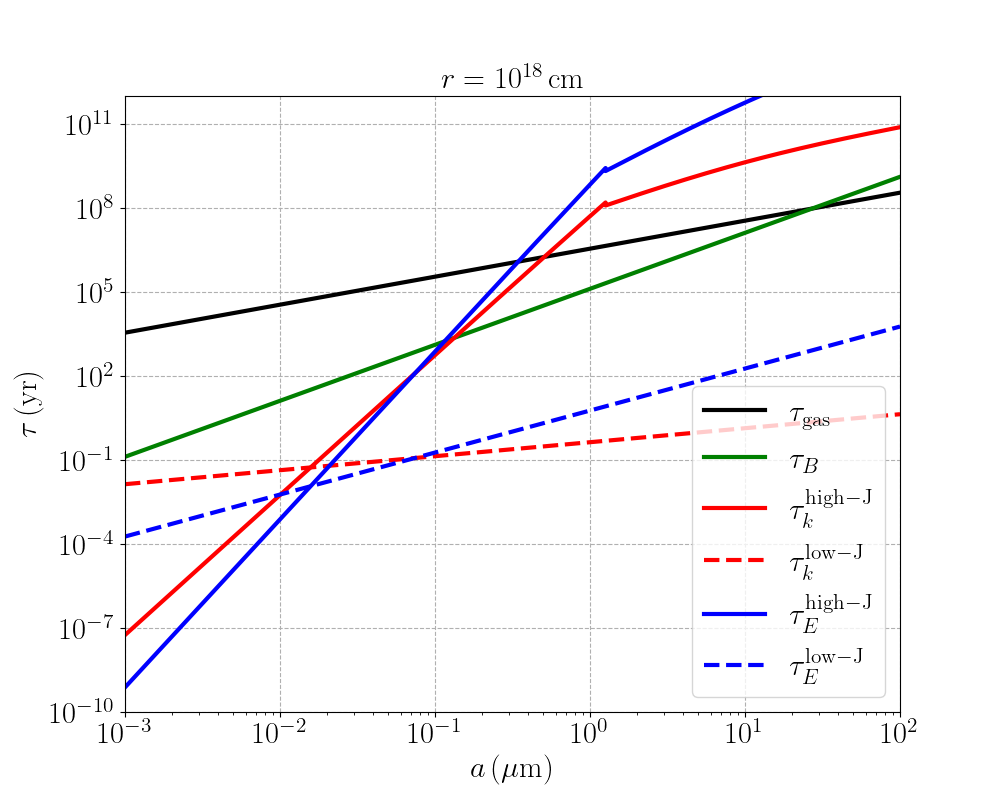}
\caption{Timescales of various processes involved in grain alignment of HAC grains at different radial distances from the center of the IRC +10216 envelope. Grains aligned at both low-J and high-J attractors are considered.}
\label{fig:timescales_AGB}
\end{figure*}

Figure \ref{fig:timescales_AGB} shows the timescales by various processes at three radii from the central star. All timescales increases with increasing the radial distances due to the decrease of the gas density and radiation strength. The intersection of the gas damping time (solid black line) with the radiative precession timescales ($\tau_{k}^{low-J,high-J}$, red solid and dashed lines) determines the critical size for grain alignment via $k$-RAT. The same applies with the alignment via $B$-RAT and $E$-RAT.

\end{document}